\documentclass{article}

\usepackage{iclr2025_conference, times}

\usepackage[utf8]{inputenc} %
\usepackage[T1]{fontenc}  %
\usepackage{hyperref}    %
\usepackage{url}      %
\usepackage{booktabs}    %
\usepackage{amsfonts}    %
\usepackage{nicefrac}    %
\usepackage{microtype}   %
\usepackage{xcolor}     %
\usepackage{graphicx}
\usepackage{float}
\usepackage{listings}
\PassOptionsToPackage{table}{xcolor}

\usepackage{multirow}
\usepackage{todonotes}
\usepackage[table]{xcolor}
\usepackage{xparse}
\usepackage[normalem]{ulem}
\NewDocumentCommand{\cross}{m g}{%
 \textcolor{orange}{\sout{#1}}%
 \IfValueT{#2}{\ \textcolor{blue}{#2}}%
}
  
\usepackage{xcolor}
\usepackage{pifont}
\usepackage{hyperref}       %
\usepackage{url}            %
\usepackage{booktabs}       %
\usepackage{amsfonts}       %
\usepackage{nicefrac}       %
\usepackage{xcolor,colortbl}         %
\usepackage{caption}
\usepackage{subcaption}
\usepackage{listings}
\usepackage{graphicx}
\usepackage{verbatim}
\usepackage{amsmath}
\usepackage[capitalize,noabbrev]{cleveref}
\usepackage{algorithm}
\usepackage{todonotes}
\usepackage{multirow}
\usepackage{placeins} 
\usepackage{amssymb}
\usepackage{svg}
\usepackage{ulem}
\usepackage{makecell}
\usepackage{pifont} 
\usepackage{xspace}
\usepackage{hyperref}
\usepackage{wrapfig}
\usepackage[at]{easylist}
\usepackage{minitoc} %
\usepackage{mathrsfs}

\noptcrule

\usepackage[toc,page,header]{appendix}
\usepackage{tikz}
\usepackage{amsmath}
\usepackage{tabularx}
\usepackage{listings}
\usepackage{array}
\usepackage{booktabs}
\usepackage{xcolor}
\usepackage{siunitx}
\usepackage{twemojis}
\usepackage[inline]{enumitem}
\usepackage[affil-it]{authblk}
\usepackage[many]{tcolorbox}
\usepackage{fancyvrb}
\usepackage{listings}
\usepackage{fvextra}
\usepackage[T1]{fontenc}
\usepackage{multirow}
\usepackage{longtable}

\definecolor{swecream}{HTML}{EFFEFF}
\definecolor{issueborder}{HTML}{15071A}
\definecolor{issuefill}{HTML}{F6F8FA}
\definecolor{envfill}{HTML}{F2F9FF}
\definecolor{envborder}{HTML}{123C7C}
\definecolor{agentfill}{HTML}{F9F3F3}
\definecolor{agentborder}{HTML}{9B0A0A}
\definecolor{goldpatchborder}{HTML}{FABB00}
\definecolor{goldpatchfill}{HTML}{FFF7E1}

\makeatletter
\newcommand{\suppressTOC}{
  \let\addcontentslineOrig\addcontentsline
  \renewcommand{\addcontentsline}[3]{}
}
\newcommand{\restoreTOC}{
  \let\addcontentsline\addcontentslineOrig
}
\makeatother

\definecolor{mydarkblue}{rgb}{0,0.08,0.55}

\definecolor{codegreen}{rgb}{0,0.6,0}
\definecolor{codegray}{rgb}{0.5,0.5,0.5}
\definecolor{codepurple}{rgb}{0.58,0,0.82}
\definecolor{backcolour}{rgb}{0.95,0.95,0.92}

\hypersetup{
    colorlinks=true,
    linkcolor=violet,
    filecolor=blue,      
    urlcolor=violet,
    citecolor=violet
}
\lstdefinestyle{mystyle}{
    backgroundcolor=\color{backcolour},   
    commentstyle=\color{codegreen},
    keywordstyle=\color{magenta},
    numberstyle=\tiny\color{codegray},
    stringstyle=\color{codepurple},
    basicstyle=\ttfamily\tiny,
    breakatwhitespace=false,         
    breaklines=true,                 
    captionpos=b,                    
    keepspaces=true,                 
    numbersep=5pt,                  
    showspaces=false,                
    showstringspaces=false,
    showtabs=false,                  
    tabsize=2,
    prebreak=\raisebox{0ex}[0ex][0ex]{\scalebox{1.5}{\color{red}\ensuremath{\hookleftarrow}}} %
}

\newcommand{\aref}[1]{\hyperref[#1]{Appendix~\ref*{#1}}}

\definecolor{t1}{rgb}{0.53, 0.73, 0.9}
\definecolor{t3}{rgb}{0.0, 0.0, 0.65}

\definecolor{lightgray}{rgb}{.9,.9,.9}
\definecolor{darkgray}{rgb}{.4,.4,.4}
\definecolor{purple}{rgb}{0.65, 0.12, 0.82}

\newcommand{\dojo}{\textsc{CTF-Dojo}\xspace}
\newcommand{\forge}{\textsc{CTF-Forge}\xspace}

\newcommand{\github}{\raisebox{-1.5pt}{\includegraphics[height=1.05em]{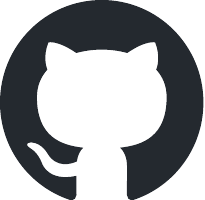}}\xspace}

\newcommand{\enigma}{\textsc{EnIGMA}\xspace}
\newcommand{\enigmaplus}{\textsc{EnIGMA+}\xspace}
\usepackage{graphicx}

\newtcolorbox{observationbox}[1][]{
        colback=envfill,
        colbacktitle=envfill,
        colframe=envborder,
        arc=5pt,
        fontupper=\small,
        fonttitle=\bfseries\color{black},
        boxrule=0.5mm,
        boxsep=1mm,
        width=\linewidth,
        breakable,
        title={Terminal Model \hfill #1},
        rounded corners,
        toptitle=0.7mm,
        bottomtitle=0.7mm
}
\newtcolorbox{goldpatchbox}[1][]{
        colback=goldpatchfill,
        colbacktitle=goldpatchfill,
        colframe=goldpatchborder,
        arc=5pt,
        fontupper=\small,
        fonttitle=\bfseries\color{black},
        boxrule=0.5mm,
        boxsep=1mm,
        width=\linewidth,
        breakable,
        title={\twemoji{1f6a9} Flag Captured \hfill #1},
        rounded corners,
        toptitle=0.7mm,
        bottomtitle=0.7mm
}
\newtcolorbox{issuebox}[1][]{
        colback=issuefill,
        colbacktitle=issuefill,
        colframe=issueborder,
        arc=5pt,
        fontupper=\small,
        fonttitle=\bfseries\color{black},
        boxrule=0.5mm,
        boxsep=1mm,
        width=\linewidth,
        breakable,
        title={CTF Challenge \hfill #1},
        rounded corners,
        toptitle=1mm
}
\newtcolorbox{agentbox}[1][]{
        colback=agentfill,
        colbacktitle=agentfill,
        colframe=agentborder,
        arc=5pt,
        fontupper=\small,
        fonttitle=\bfseries\color{black},
        boxrule=0.5mm,
        boxsep=1mm,
        width=\linewidth,
        breakable,
        title={Player Model \hfill #1},
        rounded corners,
        toptitle=1mm,
        lower separated=false
}
\newtcolorbox{fileviewerbox}[1]{
        enhanced,
        breakable,
        boxrule = 1.5pt,
        fontupper = \small,
        fonttitle = \bf\color{black},
        arc = 5pt,
        rounded corners,
        colframe = black,
        colbacktitle = swecream,
        colback = swecream,
        title = #1,
        left=4pt %
}
\newtcolorbox{promptbox}[1]{
    enhanced,
    breakable,
    boxrule=1pt,  %
    fontupper=\small,
    fonttitle=\bfseries\color{black},
    arc=3pt,  %
    rounded corners,
    colframe=black,
    colbacktitle=swecream,
    colback=swecream,
    title=#1,
    left=2mm,  %
    right=2mm,  %
    top=1mm,  %
    bottom=1mm  %
}

\DefineVerbatimEnvironment{CodeVerbatim}{Verbatim}{
  formatcom={\color{black}},
  fontsize=\small,
  fontfamily=\ttdefault,
  fontseries=\mddefault,
  fontshape=\updefault,
  fillcolor=\color{white},
  framerule=0pt,
}
\DeclareUnicodeCharacter{2501}{\textbar} 
\DeclareUnicodeCharacter{2502}{\textbar}

\newcommand{\cmark}{{\color{green!70!black}\ding{51}}} %
\newcommand{\xmark}{{\color{red}\ding{55}}}      %

\title{Training Language Model Agents to Find\\Vulnerabilities with CTF-Dojo}
\author{%
 \textbf{Terry Yue Zhuo}{$^{1,2}$}\thanks{Work done during an internship at Amazon.}\hspace{3.5mm}
 \textbf{Dingmin Wang}{$^{2}$}\hspace{2.5mm}
 \textbf{Hantian Ding}{$^{2}$}\hspace{2.5mm}
 \textbf{Varun Kumar}{$^{2}$}\hspace{2.5mm}
 \textbf{Zijian Wang}{$^{2}$}\\
$^1$\raisebox{-0.5ex}{\includegraphics[scale=0.015]{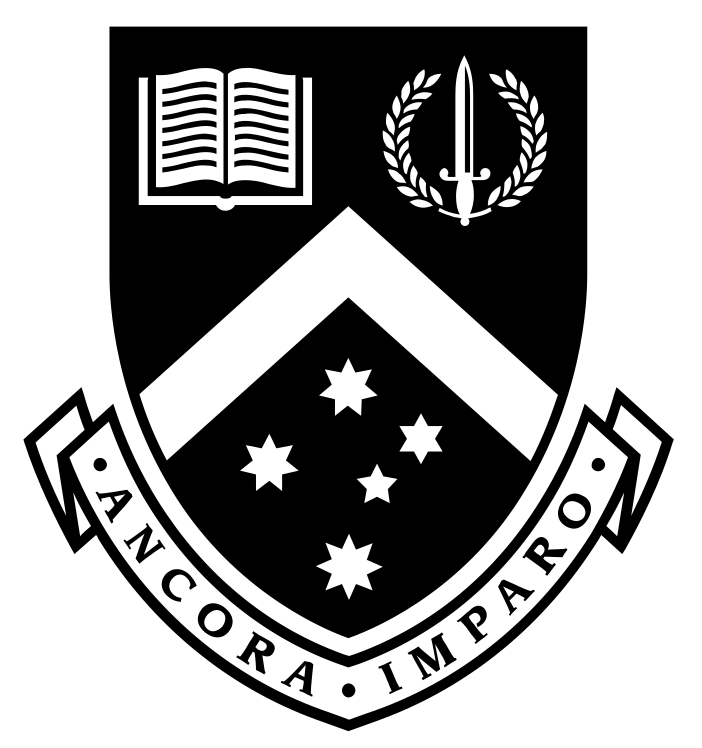}} Monash University \hspace{5mm}
$^2$\raisebox{-0.8ex}{\includegraphics[scale=0.006]{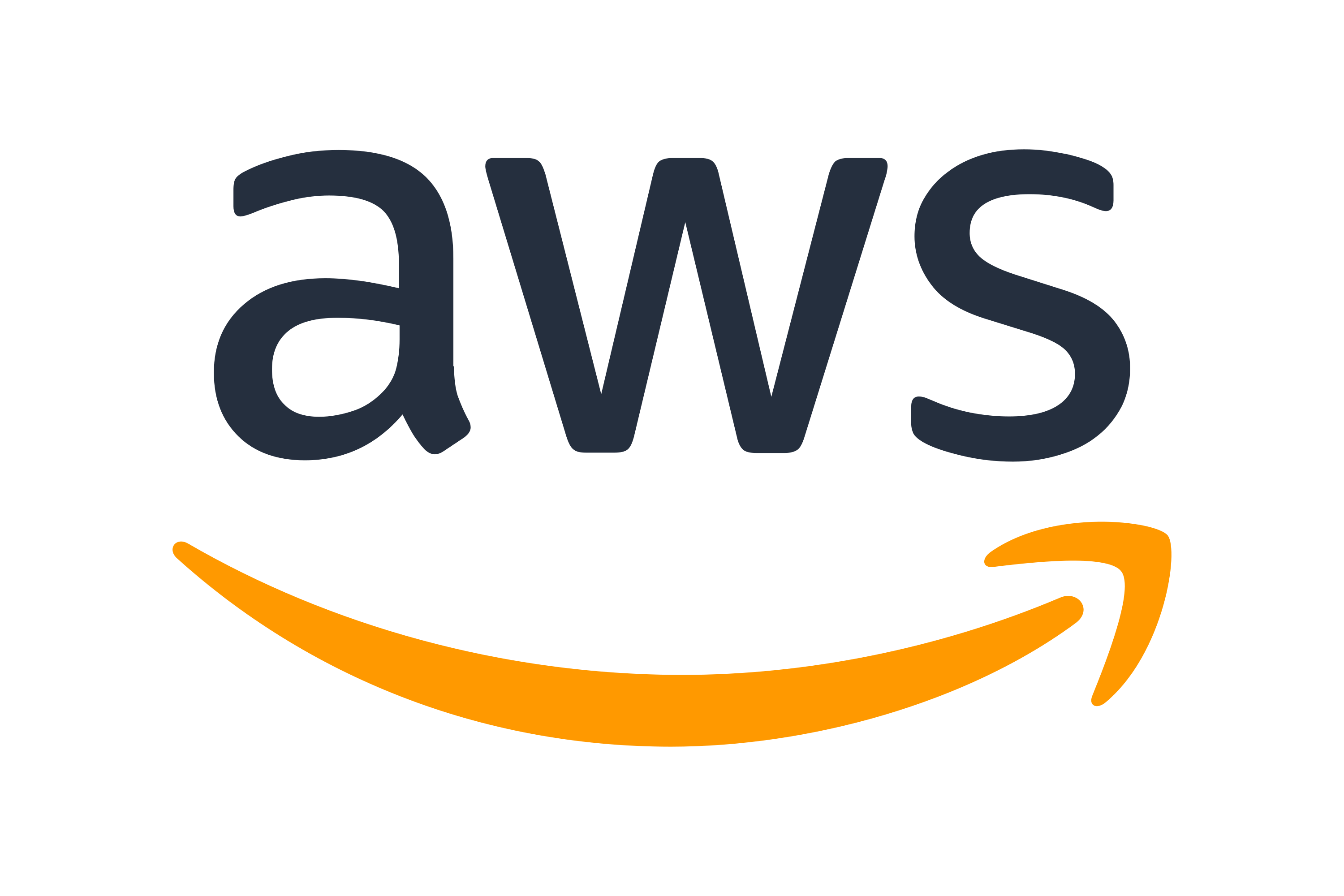}} AWS AI Labs\\
\vspace{2ex}
\texttt{terry.zhuo@monash.edu}\\
\texttt{\{wdimmy, dhantian, kuvrun, zijwan\}@amazon.com}
}
\iclrfinalcopy

\begin{document}
\maketitle
\suppressTOC

\begin{center}
\vspace{-2em}
\begin{tabular}{c}
  \github \url{https://github.com/amazon-science/CTF-Dojo} \\
\end{tabular}
\end{center}

\begin{abstract}

Large language models (LLMs) have demonstrated exceptional capabilities when trained within executable runtime environments, notably excelling at software engineering tasks through verified feedback loops. Yet, scalable and generalizable execution-grounded environments remain scarce, limiting progress in training more capable ML agents. We introduce \dojo, the first large-scale executable runtime tailored for training LLMs with verifiable feedback, featuring 658 fully functional Capture-The-Flag (CTF)-style challenges containerized in Docker with guaranteed reproducibility. To enable rapid scaling without manual intervention, we develop \forge, an automated pipeline that transforms publicly available artifacts into ready-to-use execution environments in minutes, eliminating weeks of expert configuration traditionally required.

We trained LLM-based agents on just 486 high-quality, execution-verified trajectories from \dojo, achieving up to 9.9\% absolute gains over strong baselines across three competitive benchmarks: \textit{InterCode-CTF}, \textit{NYU CTF Bench}, and \textit{Cybench}. Our best-performing 32B model reaches 31.9\% Pass@1, establishing a new open-weight state-of-the-art that rivals frontier models like DeepSeek-V3-0324 and Gemini-2.5-Flash. By framing CTF-style tasks as a benchmark for executable-agent learning, \dojo demonstrates that execution-grounded training signals are not only effective but pivotal in advancing high-performance ML agents without dependence on costly proprietary systems.

\end{abstract}

\begin{figure*}[!h]
  \centering
  \includegraphics[width=\linewidth]{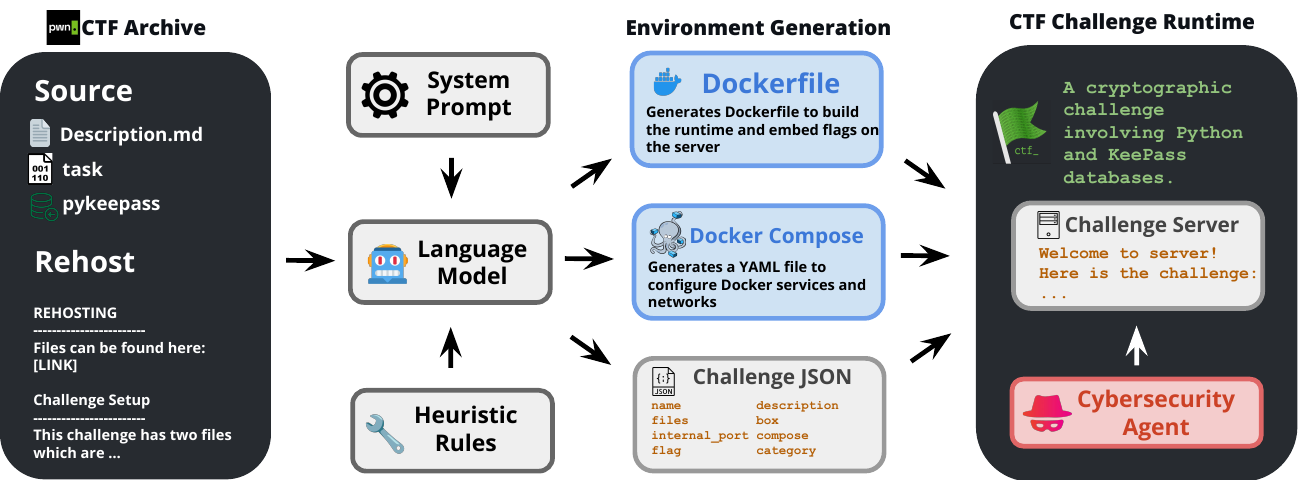}
  \caption{
  {\forge} powers automated creation of configuration files from publicly sourced CTF artifacts for containerizing CTF challenges.}
  \label{fig:pipeline}
\end{figure*}

\section{Introduction}

Advanced cybersecurity necessitates the ongoing analysis of increasingly complex software systems. As globally connected infrastructures expand, their attack surfaces expand as well, making traditional manual security analysis insufficient for timely vulnerability identification and remediation. This urgency has spurred major research efforts, such as the DARPA Cyber Grand Challenge~\citep{song2015darpa} and DARPA AIxCC~\citep{darpa2024aixcc}, which focus on building autonomous systems capable of discovering and validating software flaws. In this context, Capture The Flag (CTF) competitions have emerged as the de facto benchmark for evaluating the cybersecurity reasoning abilities of machine learning models, demanding advanced, multi-step adversarial strategies to uncover system vulnerabilities and retrieve hidden flags~\citep{anthropic2025claude37sonnet,xai2025_rmfdraft,owasp2025_llm_exploit_generation_v1}.

Previous works have demonstrated promising results in applying large language model (LLM) agents to CTF challenges~\citep{hurst2024gpt,jaech2024openai,anthropic2025claude4_opus_sonnet,abramovichenigma}, with systems like {\enigma}~\citep{abramovichenigma} achieving substantial progress on complex security tasks. While these approaches enable frontier proprietary models to achieve strong performance, they fail short when applied to open-source LLMs due to the lack of agentic training data. Recently, \citet{zhuo2025cyber} shows that training on thousands of synthetic agent trajectories can close the gap between proprietary and open-source LLMs. However, synthesizing a large number of long-horizon trajectories from teacher models requires substantial computational resources,  limiting generalization under budget constraints. Moreover, the validity of synthetic trajectories is hard to verify without runtime environments, limiting their reliability for training in high-stakes, safety-critical domains.

To address these limitations, we present \dojo, the first execution environment that contains hundreds of fully functional CTF challenges in secure Docker containers. \dojo leverages CTF artifacts (e.g., challenge descriptions and files to reproduce each challenge) from \texttt{pwn.college}, a public archive developed by Arizona State University for hands-on cybersecurity education, now used in 145 countries and actively maintained by a team of professors and students. However, setting up the runtime environment for CTF challenges is extremely difficult for non-professionals and can take up to an hour per task even for experienced practitioners (documented \autoref{sec:dojo}). To eliminate this bottleneck, we propose {\forge} (\autoref{fig:pipeline}), an automated pipeline that leverages LLMs to create hundreds of Docker images for \dojo within minutes, achieving over 98\% success rate through manual validation.

During trajectory collection from multiple LLMs within \dojo, we found that weaker models struggle to solve CTF challenges independently (detailed in \autoref{subsec:hints}). To improve yield rates, we collect diverse CTF writeups from {CTFtime}\footnote{\url{https://ctftime.org/}} and incorporated them as inference-time hints. Although we notice that only 23\% of the {\dojo} challenges matches at least one writeup, we empirically find that such writeup content, when available, can significantly boost the success rate of LLMs up to 64\% relatively gains. Notably, while building these environments, \dojo uncovered four bugs from the existing \texttt{pwn.college} collection\footnote{We have filed issues in their \href{https://github.com/pwncollege/ctf-archive}{official repository}.}.

Models trained on \dojo trajectories achieve open-weight state-of-the-art performance on over 300 tasks across three established CTF benchmarks. Through the extensive analysis, we identify three key findings for building effective cybersecurity agents: (1) writeups are crucial for training, particularly when working with data generated by weak models, (2) augmenting the runtime environment (e.g., server domains and flags) helps models yield more solved more CTF challenges, and (3) employing diverse teacher LLMs in {\dojo} leads to better task diversity and stronger performance. We hope our insights from the proposed {\dojo} can shed light on the future development of cybersecurity agents. Our work provides following contributions:

\begin{itemize}[left=0pt]
\item We introduce \dojo, the first large-scale, execution-ready environment for cybersecurity agent training, offering hundreds of verified CTF challenges in isolated Docker containers.

\item We propose {\forge}, a scalable pipeline that leverages LLMs to automate the generation of Docker-based runtime environments, achieving over 98\% success rate through manual validation.

\item We conduct thorough analysis through extensive ablation studies, identifying key factors that influence agent performance, including the presence of hint-guided trajectory collection, runtime environment augmentation, and teacher model diversity.

\end{itemize}

\section{{\dojo}: Environment for Building Powerful Cybersecurity Agents}
\label{sec:dojo}

{\dojo} is the first environment designed to synthesize verified agent trajectories for training LLMs on offensive cybersecurity tasks involving vulnerability detection and exploitation. As shown in \autoref{tab:sec_data}, existing cybersecurity execution environments either lack agentic task instance or are not designed for training purposes, creating a critical gap in the development of capable security agents. Inspired by the success of trajectory-based learning in software engineering agents~\citep{jimenezswe,yang2024swe}, \dojo adapts this paradigm to cybersecurity by sourcing publicly available CTF artifacts and transforming them into executable and interactive environments.

Different from prior pipelines for software engineering tasks~\citep{pan2024training,xie2025swe,yang2025swe}, which often require human effort or complex multi-agent systems to construct Docker environments, our approach is lightweight and fully automated. Towards that end, we introduce {\forge}, a pipeline that automatically builds Docker containers for {\dojo}. While manual setup can take up to an hour per challenge even for experts\footnote{This has been attempted by one of the authors.}, {\forge} completes each container in 0.5 seconds on average, reducing weeks of total setup time to just minutes.

\begin{table*}[!t]
\centering
\caption{{\dojo} is the first cybersecurity executable environment deriving agent trajectories for training. \textit{Detection:} whether the task requires vulnerability detection; \textit{exploitation:} whether the task needs LLMs to verify the detected vulnerabilities; \textit{Agentic:} whether each instance is repaired with an interactive environment for exploitation; \textit{Real Task:} whether each instance is developed by human experts.}

\resizebox{\linewidth}{!}{
\begin{tabular}{l|cccc|rr}
\toprule
\textbf{Executable Environment}  & \textbf{Detection} & \textbf{Exploitation} & \textbf{Agentic} & \textbf{Real Task}  & \textbf{\# Total} & \textbf{\# Train} \\
\midrule
SecRepoBench~\citep{dilgren2025secrepobench} & \xmark  & \xmark  & \cmark & \cmark & 318 & 0 \\
CVE-Bench~\citep{wang2025cve} & \xmark  & \xmark  & \cmark & \cmark & 509 & 0 \\
\midrule
CVE-Bench~\citep{zhucve} & \xmark  & \cmark  & \cmark & \cmark & 509 & 0 \\
SEC-bench~\citep{lee2025sec} & \xmark  & \cmark  & \cmark & \cmark  & 1,507 & 0 \\
CyberGym~\citep{wang2025cybergym} & \xmark  & \cmark  & \cmark & \cmark  & 1,507 & 0 \\
\midrule
CyberSecEval 3~\citep{wan2024cyberseceval} & \cmark & \cmark & \cmark & \xmark  & 6 & 0 \\
SecCodePLT~\citep{yang2024seccodeplt} & \cmark  & \cmark  & \cmark & \xmark  & 1,345 & 0 \\
\midrule
InterCode-CTF~\citep{yang2023intercode} & \cmark  & \cmark  & \cmark  & \cmark & 100 & 0\\
NYU CTF Bench~\citep{shao2024nyu}  & \cmark  & \cmark  & \cmark & \cmark  & 200 & 0\\

Cybench~\citep{zhang2025cybench} & \cmark  & \cmark & \cmark  & \cmark & 40 & 0 \\
BountyBench~\citep{zhang2025bountybench} & \cmark  & \cmark  & \cmark & \cmark   & 40 & 0\\
\rowcolor{gray!20}
{\dojo} (Ours) & \cmark  & \cmark & \cmark & \cmark  & 658 & 658 \\
\bottomrule
\end{tabular}}
\label{tab:sec_data}
\end{table*}

\subsection{Source Data Collection}

We begin by surveying CTF collections that offer diverse challenges from CTF competitions. During our initial exploration, we determine a few candidates: (1) {Sajjadium's CTF Archives}\footnote{\url{https://github.com/sajjadium/ctf-archives}}, (2) {r3kapig's Notion}\footnote{\url{https://r3kapig-not1on.notion.site}}, (3) {CryptoHack CTF Archive}\footnote{\url{https://cryptohack.org/challenges/ctf-archive/}}, (4) {archive.ooo}\footnote{\url{https://archive.ooo/}}, and (5) {\texttt{pwn.college}'s CTF Archive}\footnote{\url{https://github.com/pwncollege/ctf-archive}}. However, most of these collections suffer from inconsistent maintenance, lack standardization across challenge formats, or are limited to specific categories (e.g., CryptoHack focuses solely on cryptography). We determine that \texttt{pwn.college}'s CTF Archive is not only free of these issues but additionally provides brief information about the steps to reproduce each CTF challenge. \autoref{tab:benchmark_stats} shows the distribution of 658 CTF challenges (as of 2025/07) after decontaminating any tasks from evaluation benchmarks, 
demonstrating the diversity of CTF instances across different categories and competition events hosted between 2011 and 2025. Specially, we remove 3 CTF challenges manually as they are covered by the evaluation.

\begin{table*}[!t]
\centering
\caption{Challenge distribution across CTF datasets.}
\resizebox{\linewidth}{!}{
\begin{tabular}{l|cc|cccccc|c}
\toprule
\textbf{Benchmark} & \textbf{Level}  & \textbf{\# Competition} & \textbf{\# Crypto} & \textbf{\# Forensics} & \textbf{\# Pwn} & \textbf{\# Rev} & \textbf{\# Web} & \textbf{\# Misc} & \textbf{\# Total}\\
\midrule
\rowcolor{gray!20} \multicolumn{10}{c}{\textbf{\textit{Training}}} \\
\midrule
\dojo      & Multi-Level  & 50  & 228 & 38 & 163 & 123 & 21 & 85 & 658 \\
\midrule
\rowcolor{gray!20} \multicolumn{10}{c}{\textbf{\textit{Evaluation}}} \\
\midrule
InterCode-CTF & High School & 1 & 16  & 13 &  2 & 27 &  2 & 31 &  91 \\
NYU CTF Bench & University & 1 & 51  & 14 & 38 & 51 & 14 & 24 & 192 \\
Cybench       & Professional & 4 & 16 & 4 & 2 & 6 & 8 & 4 & 40 \\
\bottomrule
\end{tabular}}
\label{tab:benchmark_stats}
\end{table*}

CTF challenges employ two primary flag-handling mechanisms. The first type uses predefined flags, hashed with SHA-256 and verified through a provided binary executable (e.g., \texttt{flagCheck}) that confirms submission correctness. Since these flags were manually captured and encoded, they are subject to occasional errors (see 4 identified bugs in \autoref{app:bugs}). The second type relies on dynamic flag generation, where the correct flag is generated at runtime and stored in a system path such as \texttt{/flag}. In those challenges, participants must verify the system during execution to retrieve or compute the correct flag, rather than match against a static value.

\subsection{{\forge}: Automatic Environment Creation for CTF Challenges}

\autoref{fig:pipeline} illustrates \forge, a pipeline employing DeepSeek-V3-0324~\citep{liu2024deepseek} to generate environments and metadata for CTF runtime. After we source the CTF artifacts from \texttt{pwn.college}'s CTF Archive, we design a set of prompts to instruct LLMs to generate the compulsory files for Docker images in multiple stages. First, we determine whether the CTF challenge requires a containerized server to interact with. Such servers are typically needed for web challenges, binary exploitation challenges, and cryptography challenges that provide interactive services. The pipeline automatically detects server requirements by analyzing the presence of flag verification files (SHA256 checksums or check scripts) and challenge descriptions. For existing CTF runtime, we can categorize them into several challenge types: 1) Web challenges that require web servers (Apache/Nginx) to serve PHP, Python, or Node.js applications; 2) Binary exploitation challenges that need socat to host binary services on port 1337 with appropriate library dependencies; 3) Cryptography challenges that may require Python runtime environments for cryptographic services; 4) Reverse engineering challenges providing downloadable binaries and potentially analysis services; and 5) Forensics challenges offering evidence files for offline analysis. The pipeline employs category-specific guidelines and adaptive Docker setup strategies to handle different architectures (32-bit vs 64-bit), library dependencies, and runtime environments. For each challenge type, \forge generates appropriate Dockerfiles with proper base images, package installations, file copying, and service configurations, then produces \texttt{docker-compose.yml} files for orchestration and \texttt{challenge.json} metadata files that describe the challenge structure and provide flag verification mechanisms.

\subsection{Building Sustainable Environment for Cybersecurity Agents}
To ensure {\dojo} serves as a robust foundation for long-term research on autonomous cybersecurity agents, we emphasize sustainability across two dimensions: reliability and scalability.

\paragraph{Reliability} To ensure the reliability of the CTF environments created via {\forge}, we implement an automated validation script that performs two critical checks: (1) whether the Docker containers can be successfully built and executed without errors, and (2) whether the CTF services inside the containers respond correctly to network communication on the expected ports. We run {\forge} three times independently on all 658 CTF challenges to evaluate consistency and determinism. Across these runs, 98\% (650) of the challenges consistently pass all checks, demonstrating high reliability of the pipeline in producing stable, executable environments for cybersecurity agents. Additionally, we sample 10\% of the built CTF tasks and manually test the executables within each runtime to verify expected behavior.

\paragraph{Scalability} While \dojo currently contains fewer instances than existing software engineering environments that covers thousands of instances~\citep{pan2024training,xie2025swe,yang2025swe}, each CTF challenge environment is uniquely designed, mimicking diverse real-world software systems rather than variations of a single codebase that is common in SWE tasks. To enhance scalability over time, {\dojo} builds on the actively growing CTF collections from the \texttt{pwn.college} community. As new challenges are added, {\forge} can continuously and automatically convert them into interactive environments with minimal manual effort, enabling \dojo to scale organically alongside community-driven CTF development.

\subsection{Training Data Construction}
\label{subsec:data_construct}
We introduce a data pipeline to produce a large corpus of high-quality, multi-turn interaction traces from {\dojo}. This process supports the development of CTF-solving agents that require diverse, realistic demonstrations of iterative security problem-solving behavior.

\paragraph{Agent Scaffold}

We build on {\enigmaplus}~\citep{zhuo2025cyber}, a recently introduced agent scaffold designed for scalable and consistent evaluation of agents on cybersecurity tasks. {\enigmaplus} extends the original {\enigma} framework to better support cybersecurity environments by incorporating interactive tools for debugging and remote server interaction. Notably, {\enigmaplus} improves evaluation efficiency by executing tasks in parallel using isolated Docker containers, reducing runtime from days to hours for large-scale experiments. It also enables the control of agent interactions based on the number of interaction steps (e.g., 40 turns) rather than monetary cost, which aligns with best practices in agent evaluation. Additionally, it replaces \enigma's context-heavy summarization module with a lightweight alternative better suited for binary analysis outputs. Within this scaffold, we integrate the {\dojo} environment and collect agent trajectories through structured interactions.

\paragraph{Trajectory Collection}

Within the \enigmaplus scaffold, we deploy DeepSeek-V3-0324 to attempt solving CTF challenges in {\dojo} with a temperature of 0.6, top-p of 0.95, and rollout count of 6. For each challenge instance, the agent is given the original task description and interactive access to the containerized environment, capped at 40 turns. We log every system command, intermediate output, and reasoning step until either the flag is captured or the turn budget is exhausted. Successful trajectories are stored in structured JSON format for downstream filtering and training.
Our initial large-scale runs reveal that many trajectories stall due to brittle exploitation strategies or failure to discover the correct toolchain. While some challenges yield multiple successful runs, a large fraction remain unsolved or are solved only rarely, leading to a skewed dataset concentrated on limited tasks.

\paragraph{Inference-Time Bag of Tricks}

To increase the yield rate of successful trajectories on CTF challenges, we introduce two inference-time techniques (analyzed in \autoref{sec:ablation}). \textit{First, we leverage publicly available CTF writeups to provide task-specific hints to LLMs}. Specifically, we collect 8,361 writeups and apply fuzzy matching to align them with challenges in {\dojo}. This yields 252 matched writeups, covering 150 challenges with at least one relevant writeup. During preprocessing, we redact any potential flag values from the writeups and incorporate the cleaned content into the task prompt, as the direct answers may lead to the shortcut learning~\citep{geirhos2020shortcut}. Furthermore, two of the authors carefully inspected the matched writeups to confirm that no leaked flags were present and that all writeups corresponded correctly to the CTF challenges. We explicitly instruct the LLM to treat the writeup as a source of inspiration, using its strategies and reasoning implicitly without direct referencing. To ensure the integrity of downstream evaluation, we remove all writeup content from collected trajectories after inference.
In addition, to further guarantee that no residual writeup information remains, we randomly sample 20\% of the trajectories after this removal step and have two of the authors carefully verify that the agent’s reasoning does not reproduce or paraphrase the hint text. This double-check helps confirm that the final data reflect genuinely self-directed problem-solving rather than implicit reuse of the provided hints.
\textit{Second, we augment the CTF runtime per agent rollout via {\forge} by introducing randomized environment configurations}. These augmentations include varying port numbers, modifying file system paths, injecting non-functional distractor code, and adjusting system-level metadata such as timestamps and installed packages. While preserving the core logic and solvability of each challenge, these perturbations reduce overfitting to static runtime cues and encourage agents to develop more generalizable exploitation strategies. They also help mitigate persistent misconfigurations introduced by LLMs. By resetting the runtime with diverse settings, the environment is more likely to land in a valid configuration that enables flag discovery, even if previous runs failed due to deterministic setup errors. For challenges with dynamic flag generation, we re-seed the container environments at each rollout to ensure unique flag instances per interaction, further enriching training data diversity.

\begin{wrapfigure}{r}{0.3\textwidth}
  \centering
  \vspace{-35pt}
  \includegraphics[width=0.3\textwidth]{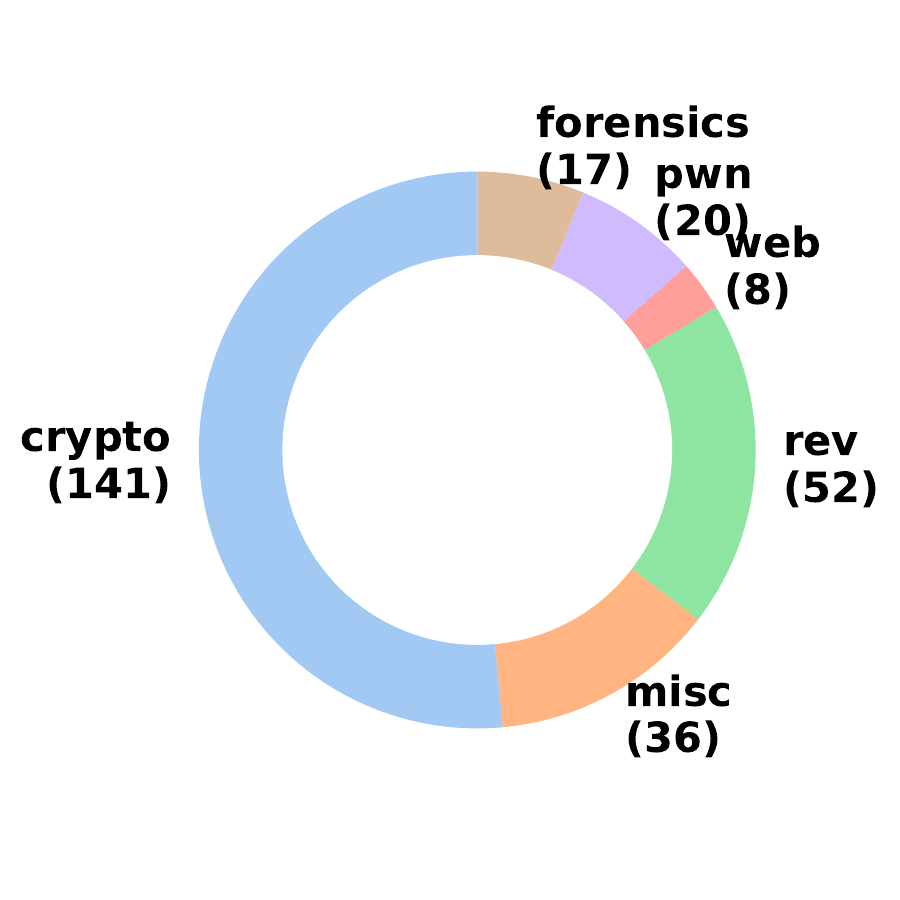}
  \vspace{-30pt}
  \caption{Solved challenges.}
  \label{fig:category_distribution}
\end{wrapfigure}

\paragraph{Data Analysis}

We employ two models, Qwen3-Coder~\citep{yang2025qwen3} and DeepSeek-V3-0324~\citep{liu2024deepseek}, to analyze the composition and characteristics of the raw 1,006 successful trajectories across multiple runs to better understand the coverage and difficulty distribution within {\dojo}. \autoref{fig:category_distribution} shows the category distribution across solved 274 challenges, where cryptography tasks constitute the largest portion, followed by reverse engineering, and miscellaneous categories. This distribution reflects the typical emphasis in modern CTFs on cryptographic reasoning and binary analysis. We provide more data analysis in \autoref{app:analysis}.

\section{Training LLMs as Cybersecurity Agents with {\dojo}}
With {\dojo}, we train cybersecurity agents with various base models. Our primary objective is to establish strong baselines and demonstrate the effectiveness of training data derived from execution. We use Pass@$k$~\citep{chen2021evaluating} as our main evaluation metric. Similar to \citet{pan2024training}, we employ a simple policy improvement algorithm: rejection sampling fine-tuning, where we fine-tune the model on trajectories successfully capturing flags inside {\dojo}. In addition, we apply sample capping of 2 per solved CTF challenges to avoid bias towards easy tasks, following \citet{pan2024training} and \citet{yang2025swe}. We finally collect 486 trajectories from the 274 CTF challenges solved by Qwen3-Coder and DeepSeek-V3-0324 (see \autoref{tab:stats}).

\subsection{Experiment Setup}
\label{sec:setup}
\paragraph{Training} We fine-tuned Qwen3 models at three scales: 8B, 14B, and 32B~\citep{yang2025qwen3}. All models undergo supervised fine-tuning on A100 GPUs via NVIDIA NeMo framework~\citep{kuchaiev2019nemo}. 
Due to computational constraints, we only retain synthesized samples within 32,768 tokens, resulting in 486 trajectories. The hyperparameters are consistently set as the global batch size of 16, the learning rate of 5e-6, and the epoch of 2.

\begin{table*}[!ht]
\caption{Pass@1 performance on benchmark tasks.  The improvements of \textbf{{\dojo}} are absolute in comparison with the Qwen3 model of corresponding sizes.} %
\centering
\resizebox{\linewidth}{!}{
\begin{tabular}{l|c|lll|l}
\toprule
\textbf{Model} & \textbf{Train Size} & \textbf{InterCode-CTF} & \textbf{NYU CTF} & \textbf{Cybench} & \textbf{Average} \\
\midrule
\multicolumn{6}{c}{\cellcolor{gray!20}\textbf{\textit{Proprietary Models
}}} \\
\midrule
Claude-3.7-Sonnet~\citep{anthropic2025claude37sonnet} & - & 86.8 & 18.2 & 30.0 & 39.0\\
Claude-3.5-Sonnet~\citep{anthropic2024claude35} & - & 85.7 & 16.7 & 25.0 & 37.2 \\
Gemini-2.5-Flash~\citep{comanici2025gemini} & - & 81.3 & 14.1 & 17.5 & 33.4\\
\midrule
\rowcolor{gray!20}
\multicolumn{6}{c}{\textbf{\textit{Open Weight Models}}} \\
\midrule
DeepSeek-V3-0324~\citep{liu2024deepseek} & - & 82.5 & 6.2 &27.5 & 30.3 \\
Kimi-K2~\citep{team2025kimi} & - & 72.5 & 4.7 & 15.0 & 25.1 \\
Qwen3-Coder~\citep{yang2025qwen3} & - & 70.3 & 5.7 & 10.0 & 24.5\\
Qwen2.5-Coder-7B-Instruct~\citep{hui2024qwen2} & - & 34.1 & 2.0 & 0.0 & 10.8 \\
Qwen2.5-Coder-14B-Instruct~\citep{hui2024qwen2} & - & 44.0 & 3.1 & 5.0 & 14.9 \\
Qwen2.5-Coder-32B-Instruct~\citep{hui2024qwen2} & - & 68.1 & 4.7 & 10.0 & 23.2 \\
Qwen3-8B~\citep{yang2025qwen3} & - & 41.8 & 1.6 & 5.0 & 13.3 \\
Qwen3-14B~\citep{yang2025qwen3} & - & 60.4 & 2.6 & 12.5 & 20.1 \\
Qwen3-32B~\citep{yang2025qwen3} & - & 65.9 & 4.7 & 5.0 & 22.0 \\ 
\midrule
{Cyber-Zero}-8B$^{*}$ \citep{zhuo2025cyber} & 9,464 & 64.8 & 6.3 & 10.0 & 23.2 \\
{Cyber-Zero}-14B$^{*}$ \citep{zhuo2025cyber} & 9,464 & 73.6 & 9.9 & 20.0 & 29.1 \\
{Cyber-Zero}-32B$^{*}$ \citep{zhuo2025cyber} & 9,464 & 82.4 & 13.5 & 17.5 & 33.4 \\
\midrule
{\dojo}-8B (Ours) & 486 & 53.8 (\textcolor{codegreen}{12.0\% $\uparrow$}) & 4.2 (\textcolor{codegreen}{2.6\% $\uparrow$})& 10.0 (\textcolor{codegreen}{5.0\% $\uparrow$})& 18.9 (\textcolor{codegreen}{5.6\% $\uparrow$})\\

{\dojo}-14B (Ours)& 486& 71.4 (\textcolor{codegreen}{11.0\% $\uparrow$})& 5.7 (\textcolor{codegreen}{3.1\% $\uparrow$})& 17.5 (\textcolor{codegreen}{5.0\% $\uparrow$})& 25.7 (\textcolor{codegreen}{5.6\% $\uparrow$})\\

{\dojo}-32B (Ours)& 486& 83.5 (\textcolor{codegreen}{17.6\% $\uparrow$})& 10.4 (\textcolor{codegreen}{5.7\% $\uparrow$})& 17.5 (\textcolor{codegreen}{12.5\% $\uparrow$})& 31.9 (\textcolor{codegreen}{9.9\% $\uparrow$}) \\
\bottomrule
\end{tabular}}
\label{tab:main_res}
\end{table*}

\paragraph{Evaluation Scaffolding} We use {\enigmaplus}, an enhanced version of the {\enigma} scaffold with several key improvements for large-scale cybersecurity evaluation. {\enigmaplus} executes evaluation tasks in parallel, significantly improving efficiency. Following \citet{zhuo2025cyber}, we cap each rollout at 40 interaction turns, replacing {\enigma}’s cost-based budget~\citep{yang2024swe} to ensure consistent evaluation across models. We also adopt the \textit{Simple Summarizer} to prevent context overflows from verbose outputs like binary decompilation.

\paragraph{Test Benchmarks} We evaluate agents on three established CTF benchmarks detailed in \autoref{tab:benchmark_stats}: InterCode-CTF benchmark comprises 100 CTF challenges collected from picoCTF, an online educational platform for high-school rated CTF challenges. NYU CTF Benchmark contains 200 CTF challenges from CSAW competitions (2017-2023), representing university-level difficulty. Cybench benchmark includes 40 CTF challenges collected from four distinct professional competitions: HackTheBox, Sekai CTF, Glacier and HKCert (2022-2024). These benchmarks collectively span six challenge categories: Cryptography, Forensics, Binary exploitation, Reverse-Engineering, Miscellaneous, and Web.
Following \citet{zhuo2025cyber}, we exclude 9 InterCode-CTF and 8 NYU CTF Bench challenges that cannot be set up, resulting in 91, 192, and 40 evaluation tasks, respectively. For evaluation, we deploy each LLM within the agent scaffold with access to the Linux Bash terminal.

\subsection{Result Analysis}

We evaluate all LLMs with the Pass@1 metric, where we sample three rollouts per task and validate whether the model captures the correct flag. \autoref{tab:main_res} presents performance comparisons between zero-shot and fine-tuned models across all benchmarks.

\begin{wrapfigure}{r}{0.4\textwidth}
  \centering
  \includegraphics[width=\linewidth]{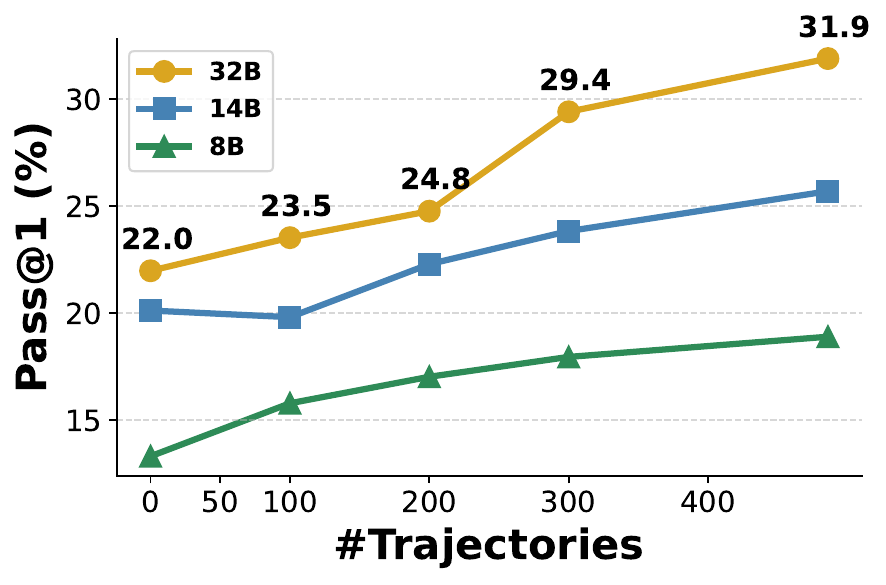}
   \caption{Effect of data scaling. Models across sizes benefit from increased number of training trajectories.}
  \label{fig:data_scaling}
\end{wrapfigure}
\paragraph{{\dojo} training enables efficient vulnerability exploitation.}
Our results show that {\dojo}-fine-tuned models achieve performance comparable to Cyber-Zero while requiring 94.9\% fewer training trajectories (486 vs. 9,464). Both approaches fine-tune on Qwen3 backbones, yet {\dojo} relies solely on a compact set of successful CTF trajectories. For instance, {\dojo}-32B reaches an average Pass@1 of 31.9\%, approaching Cyber-Zero-32B’s 33.4\%. Similarly, {\dojo}-14B achieves 25.7\% versus 29.1\% for Cyber-Zero-14B, and {\dojo}-8B attains 18.9\% compared to Cyber-Zero-8B’s 23.2\%. These results highlight that {\dojo} offers a highly data-efficient alternative: competitive performance can be attained without massive-scale training. Notably, {\dojo}-trained models also begin to rival frontier systems such as Claude-3.5-Sonnet (37.2\%), underscoring the practical feasibility of training capable cybersecurity agents at modest cost.

\paragraph{Scaling training data improves the performance linearly.}
\autoref{fig:data_scaling} shows the impact of increasing training trajectories on Pass@1 performance across different model sizes. All model variants (8B, 14B, 32B) demonstrate largely consistent performance gains as training trajectories increase, the only exception being the 14B model at 100 trajectories (19.8\% vs.\ 20.1\% zero-shot). Notably, the 32B model improves from 22.0\% to 31.9\% Pass@1 from 0 to 486 trajectories, demonstrating nearly linear performance scaling with data. This trend confirms that even modestly sized datasets can substantially enhance capability in cybersecurity tasks. Larger models not only start from higher baselines but also benefit more from additional supervision, highlighting the synergistic effect of scale and verified data in training paradigm.

\section{Ablations on {\dojo} Data}
\label{sec:ablation}
To better understand the components contributing to {\dojo}’s effectiveness, we conduct ablation studies across three axes: external writeups as inference-time hints, runtime augmentation during data collection. These experiments reveal the impact of key design choices and identify practical strategies for enhancing agent performance in cybersecurity environments. We also explore the effectiveness of teacher model diversity in \autoref{app:ablation}.

\subsection{Writups as Hints}
\label{subsec:hints}

\begin{table*}[!h]
\caption{Solved rate (\%) on {\dojo} tasks across categories, using {\enigmaplus}. ``--'' indicates baseline without writeup hints; ``+'' includes writeups in the prompt.}
\centering
\resizebox{\linewidth}{!}{
\begin{tabular}{l|cc|cc|cc|cc|cc|cc||cc}
\toprule
\multirow{2}{*}{\textbf{Models}}
& \multicolumn{2}{c|}{\textbf{\# Crypto}} 
& \multicolumn{2}{c|}{\textbf{\# Forensics}} 
& \multicolumn{2}{c|}{\textbf{\# Pwn}} 
& \multicolumn{2}{c|}{\textbf{\# Rev}} 
& \multicolumn{2}{c|}{\textbf{\# Web}} 
& \multicolumn{2}{c||}{\textbf{\# Misc}} 
& \multicolumn{2}{c}{\textbf{\# Total}} \\
\cmidrule{2-15}
& -- & + 
& -- & +  
& -- & + 
& -- & + 
& -- & + 
& -- & + 
& -- & + 
\\
\midrule
\multicolumn{15}{c}{\cellcolor{gray!20}\textbf{\textit{Proprietary Models}}} \\
\midrule
Claude-3.7-Sonnet 
& 41.2 & \textbf{50.9}
& 42.1 & \textbf{50.0} 
& 14.7 &  \textbf{20.9}
& 41.5 & \textbf{49.6} 
& 61.9 &  \textbf{76.2} 
& 47.1 &  \textbf{69.4}
& 36.2 & \textbf{46.4}
\\
Claude-3.5-Sonnet 
& 39.9 & \textbf{43.9}
& 39.5 & \textbf{47.4} 
& 8.0 &  \textbf{13.5}
& 39.8 & \textbf{41.5} 
& 47.6 &  \textbf{57.1} 
& 45.9 &  \textbf{68.2}
& 33.0 & \textbf{39.7}
\\
\midrule
\multicolumn{15}{c}{\cellcolor{gray!20}\textbf{\textit{Open Weight Models}}} \\
\midrule
DeepSeek-V3-0324

& 37.1 &  \textbf{41.0}
& 41.0 &  \textbf{43.6}
& 12.0 &  \textbf{13.5}
& 34.1 &  \textbf{36.6}
& 33.3 &  \textbf{52.4}
& 36.5 &  \textbf{41.2}
& 30.4 &  \textbf{33.9}
\\
Qwen3-Coder
& 31.4 & \textbf{42.8}
& 35.9 & \textbf{38.5}
& 7.9 & \textbf{9.1}
& 26.8 & \textbf{39.8}
& 23.8 & \textbf{28.6}
& 24.7 & \textbf{37.6}
& 23.9 & \textbf{32.5}
\\
Qwen3-32B    
& 21.9 & \textbf{29.4}
& 7.9 & \textbf{18.4} 
& 1.8 & \textbf{6.7} 
& 22.8 & \textbf{28.5} 
& 9.5 & \textbf{23.5} 
& 31.8 & \textbf{41.2} 
& 17.2 & \textbf{24.3}
\\
Qwen3-14B
& 14.0 &  \textbf{25.9}
& 5.3 & \textbf{10.5} 
& 1.8 & \textbf{4.9} 
& 20.3 & \textbf{25.2} 
& 9.5 & \textbf{14.3} 
& 24.7 & \textbf{40.0} 
& 12.9 & \textbf{21.1}
\\
\bottomrule
\end{tabular}}
\label{tab:dojo_results}
\end{table*}

\paragraph{Setup}
To assess the value of incorporating external CTF writeups during data collection, we conduct a controlled ablation on {\dojo} challenges. We compare two settings: (1) No-Hint (-), where models receive only the original challenge description, and (2) With-Hint (+), where one redacted matched writeups is randomly chosen to prepend to the prompt as a non-referential hint for the corresponding challenge. All other settings remain constant with the main experiments.

\paragraph{Analysis}
As shown in Table~\ref{tab:dojo_results}, writeup-based hints consistently improve the number of solved tasks across all models and challenge categories. On average, the number of solved challenges increases by 7.4\%, from 168 (No-Hint) to 217 (With-Hint), underscoring the utility of public writeups for improving the yield rate of training trajectories. This effect is particularly pronounced in the Crypto, Reverse Engineering, and Miscellaneous categories where solution strategies often rely on reusable heuristics or canonical exploration workflows.
This finding suggests that writeups can serve as a rich reservoir of domain-specific knowledge, allowing models to bootstrap strategic reasoning and explore more promising solution paths. We believe the effectiveness of inference-time hints can generalize to various agent tasks like solving GitHub issues~\citep{jimenezswe}, where more diverse data can be distilled from LLMs to train stronger agentic models

\subsection{Augmenting CTF Runtimes}

\begin{wrapfigure}{r}{0.45\textwidth}
    \centering
    \vspace{-2em}
    \includegraphics[width=\linewidth]{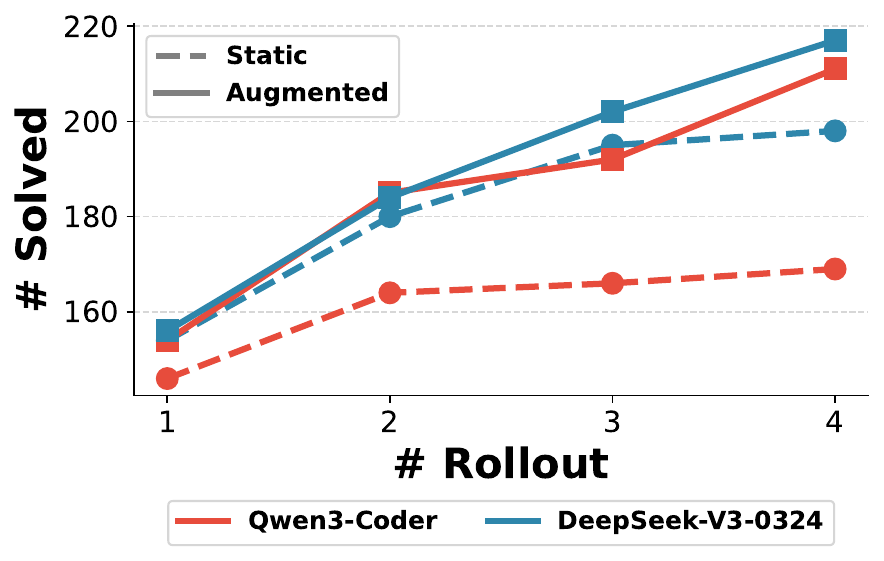}
    \caption{Effect of runtime augmentation.}

    \label{fig:augmentation}
\end{wrapfigure}
\paragraph{Setup}
To evaluate the effect of runtime augmentation on agent performance, we compare two settings for environment construction: (1) Static, where each CTF instance uses fixed runtime parameters, and (2) Augmented, where we introduce perturbations such as randomized port numbers, file path shuffling, distractor code injection, and dynamic flag regeneration. We run both Qwen3-Coder and DeepSeek-V3-0324 across 1 to 4 agent rollouts and count the number of unique CTF challenges successfully solved at least once under each setting. We keep all rollout and decoding hyperparameters identical across both variants to isolate the impact of augmentation.

\paragraph{Analysis}
\autoref{fig:augmentation} shows that augmented environments consistently yield more solved tasks across all rollout counts and both models. For example, Qwen3-Coder solves 211 challenges under augmentation at rollout 4, a relative improvement of 24.9\% compared to only 169 under static runtimes. Similarly, DeepSeek-V3-0324 improves from 156 to 217 solved tasks with augmentation at rollout 4. The performance gap widens with more rollouts, suggesting that augmentation amplifies agent exploration and generalization as more interactions are permitted. These results confirm that runtime diversity prevents brittle overfitting to environment artifacts and encourages the development of more robust, transferable strategies for flag capture.

\section{Related Work}

\paragraph{LLM Agents for Offensive Cybersecurity}

LLM agents are increasingly being applied to offensive cybersecurity, particularly in solving CTF challenges within dockerized environments~\citep{yang2023intercode,shao2024nyu,zhang2025cybench,mayoral2025cai}. These systems often build on Kali Linux due to its extensive suite of pre-installed security tools, serving as foundations for broader applications such as penetration testing, vulnerability exploitation, and cyberattack automation~\citep{charan2023text,deng2024pentestgpt,fang2024llm}. To evaluate the risks and offensive potential of such systems, benchmarks like CyberSecEval~\citep{bhatt2023purple,wan2024cyberseceval} have been proposed, while others assess the ``dangerous capabilities'' of LLMs in tasks like CTFs and red-teaming~\citep{phuong2024evaluating,guo2024redcode}, though these models still show limited performance on more complex tasks. Recent efforts have advanced agent design. Project Naptime~\citep{glazunov2024naptime} and Big Sleep~\citep{allamanis2024bigsleep} demonstrated agents capable of discovering new SQLite vulnerabilities using integrated tools like debuggers and browsers. EnIGMA~\citep{abramovichenigma} further raises the bar by combining cybersecurity-specific tools and interactive environments tailored for LLMs, achieving state-of-the-art results. Recently, \citet{zhuo2025cyber} introduced Cyber-Zero, achieving the best performance among open-source LLMs. Unlike prior methods that primarily depend on inference-time scaffolds or unverified training data, we introduce a runtime environment that efficiently enhances model performance via execution.

\paragraph{Benchmarking Models' Cybersecurity Capabilities}

Several benchmarks have been proposed to evaluate LLMs on cybersecurity tasks. Multiple-choice datasets~\citep{li2024wmdp,tihanyi2024cybermetric,liu2023secqa} offer limited insight, as their results are often highly sensitive to prompt phrasing~\citep{qi2024evaluating,lucki2024adversarial} and lack alignment with real-world operational contexts. AutoAdvExBench~\citep{carlini2025autoadvexbench} assesses LLMs' ability to autonomously break image-based adversarial defenses, while CyberSecEval~\citep{bhatt2023purple} focuses on single-turn code exploitation, capturing only a narrow slice of the interactive, multi-step nature of real-world attacks.
In contrast, agent-based frameworks with integrated tool usage offer more realistic evaluations. As a result, Capture-the-Flag (CTF) challenges have become a popular proxy for measuring security capabilities. Recent systems~\citep{abramovichenigma,mayoral2025cai} further enhance realism by combining interactive environments with structured, chain-of-exploitation evaluations.

\section{Conclusion and Future Work}

\paragraph{Conclusion}
We present {\dojo}, the first large-scale execution environment for training cybersecurity LLM agents, addressing the long-standing challenge of limited runtime support in this domain. Powered by our automated pipeline {\forge}, {\dojo} transforms public CTF artifacts into ready-to-use Docker containers in minutes, enabling scalable and reproducible trajectory collection. Training on just 486 high-quality agent trajectories synthesized through {\dojo}, our open-weight LLMs outperform strong baselines by up to 9.9\% on three major CTF benchmarks. Our 32B model achieves state-of-the-art results among open models, approaching the performance of Claude-3.5-Sonnet and DeepSeek-V3-0324. Our findings highlight the critical role of writeup-augmented training, runtime augmentations, and diverse agent behaviors in building effective cybersecurity models. Overall, {\dojo} provides a scalable and democratized foundation for advancing LLM-based security systems.

\paragraph{Future Work}
This work opens several promising avenues for research. First, we envision a live CTF benchmark where models are continuously evaluated on challenges collected from active competitions. By leveraging {\forge} to dynamically reconstruct and containerize these environments, we can enable scalable, real-time benchmarking and trajectory collection without manual engineering.
Second, while {\dojo} provides execution-verified data, it is limited by the static nature and finite scale of its current dataset (658 challenges). Exploring reinforcement learning is a natural next step, allowing agents to learn more generalizable strategies and handle novel problems via partial rewards or flag-based signals.
Finally, although we focused on the \texttt{pwn.college} CTF Archive for its standardized format and ease of containerization, {\forge} is not tied to this source. Extending to more heterogeneous CTF repositories will primarily require stronger environment-configuration strategies, for example by combining {\forge} with agentic approaches where LLMs autonomously infer dependencies and validate build setups.

\section*{Ethics Statement}

We recognize the dual-use implications of our work. While {\dojo} is intended to enhance cybersecurity by empowering developers and researchers to proactively identify and remediate vulnerabilities through automated penetration testing, the same techniques could also be misused for offensive purposes, such as discovering vulnerabilities in external systems or crafting malicious exploits. The nature of our approach further heightens this concern by lowering the technical barrier to training powerful cybersecurity agents.

Our results show that models trained on {\dojo}-generated trajectories can reach performance levels comparable to leading proprietary systems, underscoring that the democratization of advanced cybersecurity capabilities is not only possible but imminent. As LLM-based security tools become more capable, we emphasize the need for sustained collaboration among researchers, developers, and safety organizations to guide their responsible development and use. We believe that open research, paired with thoughtful safeguards, remains essential for ensuring these technologies ultimately strengthen cybersecurity defenses.

\section*{Reproducibility Statement}
All implementation details, including environment configuration and hyperparameter settings, are provided in \autoref{subsec:data_construct}. The evaluation setup and the procedure for generating multiple trajectories are described in \autoref{sec:setup}. To support open-science research, we release the complete source code and data-processing pipeline under an open-source license upon publication.

Due to restrictions on using proprietary frontier models for data distillation during training, we avoid any models from the organizations like OpenAI and Anthropic. To ensure both reproducibility and cost efficiency, all experiments are conducted with DeepSeek-V3-0324.

\section*{Acknowledgement}
We are deeply grateful to the {\enigma} team for open-sourcing the agent scaffold and reformatted benchmark data. We thank Yangruibo Ding for valuable early discussions and the \texttt{pwn.college} team (e.g., Yan Shoshitaishvili and Pratham Gupta) for initiating and maintaining one of the largest CTF archives that collects hundreds of verified CTF competitions. In addition, we thank Anoop Deoras and Stefano Soatto for their support. Lastly, we would like to address our appreciation to every CTF player taking the time to write detailed, informative writeups contributing to the collective knowledge that makes research like ours possible.

\clearpage
\bibliographystyle{iclr2026_conference}
\bibliography{reference}

\begin{thebibliography}{58}
\providecommand{\natexlab}[1]{#1}
\providecommand{\url}[1]{\texttt{#1}}
\expandafter\ifx\csname urlstyle\endcsname\relax
  \providecommand{\doi}[1]{doi: #1}\else
  \providecommand{\doi}{doi: \begingroup \urlstyle{rm}\Url}\fi

\bibitem[Abramovich et~al.(2025)Abramovich, Udeshi, Shao, Lieret, Xi, Milner, Jancheska, Yang, Jimenez, Khorrami, et~al.]{abramovichenigma}
Talor Abramovich, Meet Udeshi, Minghao Shao, Kilian Lieret, Haoran Xi, Kimberly Milner, Sofija Jancheska, John Yang, Carlos~E Jimenez, Farshad Khorrami, et~al.
\newblock Enigma: Interactive tools substantially assist lm agents in finding security vulnerabilities.
\newblock In \emph{Forty-second International Conference on Machine Learning}, 2025.

\bibitem[Allamanis et~al.(2024)Allamanis, Arjovsky, Blundell, Buesing, Brand, Glazunov, Maier, Maniatis, Marinho, Michalewski, Sen, Sutton, Tulsyan, Vanotti, Weber, and Zheng]{allamanis2024bigsleep}
Miltiadis Allamanis, Martin Arjovsky, Charles Blundell, Lars Buesing, Maddie Brand, Sergei Glazunov, David Maier, Petros Maniatis, Guilherme Marinho, Henryk Michalewski, Koushik Sen, Charles Sutton, Varun Tulsyan, Matteo Vanotti, Thomas Weber, and Dawn Zheng.
\newblock From naptime to big sleep: Using large language models to catch vulnerabilities in real-world code.
\newblock \url{https://googleprojectzero.blogspot.com/2024/10/from-naptime-to-big-sleep.html}, November 2024.
\newblock Accessed July 2025.

\bibitem[{Anthropic}(2024)]{anthropic2024claude35}
{Anthropic}.
\newblock {Claude 3.5 Model Card Addendum}.
\newblock \url{https://www-cdn.anthropic.com/fed9cc193a14b84131812372d8d5857f8f304c52/Model_Card_Claude_3_Addendum.pdf}, 2024.

\bibitem[{Anthropic}(2025{\natexlab{a}})]{anthropic2025claude37sonnet}
{Anthropic}.
\newblock {Claude 3.7 “Sonnet” System Card}.
\newblock \url{https://assets.anthropic.com/m/785e231869ea8b3b/original/claude-3-7-sonnet-system-card.pdf}, 2025{\natexlab{a}}.

\bibitem[{Anthropic}(2025{\natexlab{b}})]{anthropic2025claude4_opus_sonnet}
{Anthropic}.
\newblock {System Card: Claude Opus 4 \& Claude Sonnet 4}.
\newblock Technical report, Anthropic, May 2025{\natexlab{b}}.

\bibitem[Bhatt et~al.(2023)Bhatt, Chennabasappa, Nikolaidis, Wan, Evtimov, Gabi, Song, Ahmad, Aschermann, Fontana, et~al.]{bhatt2023purple}
Manish Bhatt, Sahana Chennabasappa, Cyrus Nikolaidis, Shengye Wan, Ivan Evtimov, Dominik Gabi, Daniel Song, Faizan Ahmad, Cornelius Aschermann, Lorenzo Fontana, et~al.
\newblock Purple llama cyberseceval: A secure coding benchmark for language models.
\newblock \emph{arXiv preprint arXiv:2312.04724}, 2023.

\bibitem[Carlini et~al.(2025)Carlini, Rando, Debenedetti, Nasr, and Tram{\`e}r]{carlini2025autoadvexbench}
Nicholas Carlini, Javier Rando, Edoardo Debenedetti, Milad Nasr, and Florian Tram{\`e}r.
\newblock Autoadvexbench: Benchmarking autonomous exploitation of adversarial example defenses.
\newblock \emph{arXiv preprint arXiv:2503.01811}, 2025.

\bibitem[Charan et~al.(2023)Charan, Chunduri, Anand, and Shukla]{charan2023text}
PV~Charan, Hrushikesh Chunduri, P~Mohan Anand, and Sandeep~K Shukla.
\newblock From text to mitre techniques: Exploring the malicious use of large language models for generating cyber attack payloads.
\newblock \emph{arXiv preprint arXiv:2305.15336}, 2023.

\bibitem[Chen et~al.(2021)Chen, Tworek, Jun, Yuan, Pinto, Kaplan, Edwards, Burda, Joseph, Brockman, et~al.]{chen2021evaluating}
Mark Chen, Jerry Tworek, Heewoo Jun, Qiming Yuan, Henrique Ponde De~Oliveira Pinto, Jared Kaplan, Harri Edwards, Yuri Burda, Nicholas Joseph, Greg Brockman, et~al.
\newblock Evaluating large language models trained on code.
\newblock \emph{arXiv preprint arXiv:2107.03374}, 2021.

\bibitem[Comanici et~al.(2025)Comanici, Bieber, Schaekermann, Pasupat, Sachdeva, Dhillon, Blistein, Ram, Zhang, Rosen, et~al.]{comanici2025gemini}
Gheorghe Comanici, Eric Bieber, Mike Schaekermann, Ice Pasupat, Noveen Sachdeva, Inderjit Dhillon, Marcel Blistein, Ori Ram, Dan Zhang, Evan Rosen, et~al.
\newblock Gemini 2.5: Pushing the frontier with advanced reasoning, multimodality, long context, and next generation agentic capabilities.
\newblock \emph{arXiv preprint arXiv:2507.06261}, 2025.

\bibitem[{DARPA}(2024)]{darpa2024aixcc}
{DARPA}.
\newblock {DARPA AIxCC, 2024}.
\newblock \url{https://aicyberchallenge.com/about/}, 2024.

\bibitem[Deng et~al.(2024)Deng, Liu, Mayoral-Vilches, Liu, Li, Xu, Zhang, Liu, Pinzger, and Rass]{deng2024pentestgpt}
Gelei Deng, Yi~Liu, V{\'\i}ctor Mayoral-Vilches, Peng Liu, Yuekang Li, Yuan Xu, Tianwei Zhang, Yang Liu, Martin Pinzger, and Stefan Rass.
\newblock $\{$PentestGPT$\}$: Evaluating and harnessing large language models for automated penetration testing.
\newblock In \emph{33rd USENIX Security Symposium (USENIX Security 24)}, pp.\  847--864, 2024.

\bibitem[Dilgren et~al.(2025)Dilgren, Chiniya, Griffith, Ding, and Chen]{dilgren2025secrepobench}
Connor Dilgren, Purva Chiniya, Luke Griffith, Yu~Ding, and Yizheng Chen.
\newblock Secrepobench: Benchmarking llms for secure code generation in real-world repositories.
\newblock \emph{arXiv preprint arXiv:2504.21205}, 2025.

\bibitem[Fang et~al.(2024)Fang, Bindu, Gupta, Zhan, and Kang]{fang2024llm}
Richard Fang, Rohan Bindu, Akul Gupta, Qiusi Zhan, and Daniel Kang.
\newblock Llm agents can autonomously hack websites.
\newblock \emph{arXiv preprint arXiv:2402.06664}, 2024.

\bibitem[Geirhos et~al.(2020)Geirhos, Jacobsen, Michaelis, Zemel, Brendel, Bethge, and Wichmann]{geirhos2020shortcut}
Robert Geirhos, J{\"o}rn-Henrik Jacobsen, Claudio Michaelis, Richard Zemel, Wieland Brendel, Matthias Bethge, and Felix~A Wichmann.
\newblock Shortcut learning in deep neural networks.
\newblock \emph{Nature Machine Intelligence}, 2\penalty0 (11):\penalty0 665--673, 2020.

\bibitem[Glazunov \& Brand(2024)Glazunov and Brand]{glazunov2024naptime}
Sergei Glazunov and Maddie Brand.
\newblock Project naptime: Evaluating offensive security capabilities of large language models.
\newblock \url{https://googleprojectzero.blogspot.com/2024/06/project-naptime.html}, June 2024.
\newblock Accessed July 2025.

\bibitem[Grattafiori et~al.(2024)Grattafiori, Dubey, Jauhri, Pandey, Kadian, Al-Dahle, Letman, Mathur, Schelten, Vaughan, et~al.]{grattafiori2024llama}
Aaron Grattafiori, Abhimanyu Dubey, Abhinav Jauhri, Abhinav Pandey, Abhishek Kadian, Ahmad Al-Dahle, Aiesha Letman, Akhil Mathur, Alan Schelten, Alex Vaughan, et~al.
\newblock The llama 3 herd of models.
\newblock \emph{arXiv preprint arXiv:2407.21783}, 2024.

\bibitem[Guo et~al.(2024)Guo, Liu, Xie, Zhou, Zeng, Lin, Song, and Li]{guo2024redcode}
Chengquan Guo, Xun Liu, Chulin Xie, Andy Zhou, Yi~Zeng, Zinan Lin, Dawn Song, and Bo~Li.
\newblock Redcode: Risky code execution and generation benchmark for code agents.
\newblock \emph{Advances in Neural Information Processing Systems}, 37:\penalty0 106190--106236, 2024.

\bibitem[Hui et~al.(2024)Hui, Yang, Cui, Yang, Liu, Zhang, Liu, Zhang, Yu, Lu, et~al.]{hui2024qwen2}
Binyuan Hui, Jian Yang, Zeyu Cui, Jiaxi Yang, Dayiheng Liu, Lei Zhang, Tianyu Liu, Jiajun Zhang, Bowen Yu, Keming Lu, et~al.
\newblock Qwen2. 5-coder technical report.
\newblock \emph{arXiv preprint arXiv:2409.12186}, 2024.

\bibitem[Hurst et~al.(2024)Hurst, Lerer, Goucher, Perelman, Ramesh, Clark, Ostrow, Welihinda, Hayes, Radford, et~al.]{hurst2024gpt}
Aaron Hurst, Adam Lerer, Adam~P Goucher, Adam Perelman, Aditya Ramesh, Aidan Clark, AJ~Ostrow, Akila Welihinda, Alan Hayes, Alec Radford, et~al.
\newblock Gpt-4o system card.
\newblock \emph{arXiv preprint arXiv:2410.21276}, 2024.

\bibitem[Jaech et~al.(2024)Jaech, Kalai, Lerer, Richardson, El-Kishky, Low, Helyar, Madry, Beutel, Carney, et~al.]{jaech2024openai}
Aaron Jaech, Adam Kalai, Adam Lerer, Adam Richardson, Ahmed El-Kishky, Aiden Low, Alec Helyar, Aleksander Madry, Alex Beutel, Alex Carney, et~al.
\newblock Openai o1 system card.
\newblock \emph{arXiv preprint arXiv:2412.16720}, 2024.

\bibitem[Jimenez et~al.(2024)Jimenez, Yang, Wettig, Yao, Pei, Press, and Narasimhan]{jimenezswe}
Carlos~E Jimenez, John Yang, Alexander Wettig, Shunyu Yao, Kexin Pei, Ofir Press, and Karthik~R Narasimhan.
\newblock Swe-bench: Can language models resolve real-world github issues?
\newblock In \emph{The Twelfth International Conference on Learning Representations}, 2024.

\bibitem[Kuchaiev et~al.(2019)Kuchaiev, Li, Nguyen, Hrinchuk, Leary, Ginsburg, Kriman, Beliaev, Lavrukhin, Cook, et~al.]{kuchaiev2019nemo}
Oleksii Kuchaiev, Jason Li, Huyen Nguyen, Oleksii Hrinchuk, Ryan Leary, Boris Ginsburg, Samuel Kriman, Stanislav Beliaev, Vitaly Lavrukhin, Jack Cook, et~al.
\newblock Nemo: a toolkit for building ai applications using neural modules.
\newblock \emph{arXiv preprint arXiv:1909.09577}, 2019.

\bibitem[Lee et~al.(2025)Lee, Zhang, Lu, and Zhang]{lee2025sec}
Hwiwon Lee, Ziqi Zhang, Hanxiao Lu, and Lingming Zhang.
\newblock Sec-bench: Automated benchmarking of llm agents on real-world software security tasks.
\newblock \emph{arXiv preprint arXiv:2506.11791}, 2025.

\bibitem[Li et~al.(2024)Li, Pan, Gopal, Yue, Berrios, Gatti, Li, Dombrowski, Goel, Mukobi, et~al.]{li2024wmdp}
Nathaniel Li, Alexander Pan, Anjali Gopal, Summer Yue, Daniel Berrios, Alice Gatti, Justin~D Li, Ann-Kathrin Dombrowski, Shashwat Goel, Gabriel Mukobi, et~al.
\newblock The wmdp benchmark: measuring and reducing malicious use with unlearning.
\newblock In \emph{Proceedings of the 41st International Conference on Machine Learning}, pp.\  28525--28550, 2024.

\bibitem[Li et~al.(2023)Li, Allal, Zi, Muennighoff, Kocetkov, Mou, Marone, Akiki, Li, Chim, et~al.]{li2023starcoder}
Raymond Li, Loubna~Ben Allal, Yangtian Zi, Niklas Muennighoff, Denis Kocetkov, Chenghao Mou, Marc Marone, Christopher Akiki, Jia Li, Jenny Chim, et~al.
\newblock Starcoder: may the source be with you!
\newblock \emph{arXiv preprint arXiv:2305.06161}, 2023.

\bibitem[Liu et~al.(2024)Liu, Feng, Xue, Wang, Wu, Lu, Zhao, Deng, Zhang, Ruan, et~al.]{liu2024deepseek}
Aixin Liu, Bei Feng, Bing Xue, Bingxuan Wang, Bochao Wu, Chengda Lu, Chenggang Zhao, Chengqi Deng, Chenyu Zhang, Chong Ruan, et~al.
\newblock Deepseek-v3 technical report.
\newblock \emph{arXiv preprint arXiv:2412.19437}, 2024.

\bibitem[Liu(2023)]{liu2023secqa}
Zefang Liu.
\newblock Secqa: A concise question-answering dataset for evaluating large language models in computer security.
\newblock \emph{arXiv preprint arXiv:2312.15838}, 2023.

\bibitem[Lozhkov et~al.(2024)Lozhkov, Li, Allal, Cassano, Lamy-Poirier, Tazi, Tang, Pykhtar, Liu, Wei, et~al.]{lozhkov2024starcoder}
Anton Lozhkov, Raymond Li, Loubna~Ben Allal, Federico Cassano, Joel Lamy-Poirier, Nouamane Tazi, Ao~Tang, Dmytro Pykhtar, Jiawei Liu, Yuxiang Wei, et~al.
\newblock Starcoder 2 and the stack v2: The next generation.
\newblock \emph{arXiv preprint arXiv:2402.19173}, 2024.

\bibitem[{\L}ucki et~al.(2024){\L}ucki, Wei, Huang, Henderson, Tram{\`e}r, and Rando]{lucki2024adversarial}
Jakub {\L}ucki, Boyi Wei, Yangsibo Huang, Peter Henderson, Florian Tram{\`e}r, and Javier Rando.
\newblock An adversarial perspective on machine unlearning for ai safety.
\newblock \emph{arXiv preprint arXiv:2409.18025}, 2024.

\bibitem[Ma et~al.(2024)Ma, Cao, Cao, Zhang, Chen, Liu, Liu, Li, Huang, and Li]{ma2024lingma}
Yingwei Ma, Rongyu Cao, Yongchang Cao, Yue Zhang, Jue Chen, Yibo Liu, Yuchen Liu, Binhua Li, Fei Huang, and Yongbin Li.
\newblock Lingma swe-gpt: An open development-process-centric language model for automated software improvement.
\newblock \emph{arXiv preprint arXiv:2411.00622}, 2024.

\bibitem[Mayoral-Vilches et~al.(2025)Mayoral-Vilches, Navarrete-Lozano, Sanz-G{\'o}mez, Espejo, Crespo-{\'A}lvarez, Oca-Gonzalez, Balassone, Glera-Pic{\'o}n, Ayucar-Carbajo, Ruiz-Alcalde, et~al.]{mayoral2025cai}
V{\'\i}ctor Mayoral-Vilches, Luis~Javier Navarrete-Lozano, Mar{\'\i}a Sanz-G{\'o}mez, Lidia~Salas Espejo, Marti{\~n}o Crespo-{\'A}lvarez, Francisco Oca-Gonzalez, Francesco Balassone, Alfonso Glera-Pic{\'o}n, Unai Ayucar-Carbajo, Jon~Ander Ruiz-Alcalde, et~al.
\newblock Cai: An open, bug bounty-ready cybersecurity ai.
\newblock \emph{arXiv preprint arXiv:2504.06017}, 2025.

\bibitem[Muennighoff et~al.(2024)Muennighoff, Liu, Zebaze, Zheng, Hui, Zhuo, Singh, Tang, Von~Werra, and Longpre]{muennighoffoctopack}
Niklas Muennighoff, Qian Liu, Armel~Randy Zebaze, Qinkai Zheng, Binyuan Hui, Terry~Yue Zhuo, Swayam Singh, Xiangru Tang, Leandro Von~Werra, and Shayne Longpre.
\newblock Octopack: Instruction tuning code large language models.
\newblock In \emph{The Twelfth International Conference on Learning Representations}, 2024.

\bibitem[{OWASP GenAI Project (CTI Layer Team)}(2025)]{owasp2025_llm_exploit_generation_v1}
{OWASP GenAI Project (CTI Layer Team)}.
\newblock {OWASP LLM Exploit Generation Version 1.0}.
\newblock Technical report, OWASP GenAI Project, February 2025.
\newblock Accessed: 3 July 2025.

\bibitem[Pan et~al.(2024)Pan, Wang, Neubig, Jaitly, Ji, Suhr, and Zhang]{pan2024training}
Jiayi Pan, Xingyao Wang, Graham Neubig, Navdeep Jaitly, Heng Ji, Alane Suhr, and Yizhe Zhang.
\newblock Training software engineering agents and verifiers with swe-gym.
\newblock \emph{arXiv preprint arXiv:2412.21139}, 2024.

\bibitem[Phuong et~al.(2024)Phuong, Aitchison, Catt, Cogan, Kaskasoli, Krakovna, Lindner, Rahtz, Assael, Hodkinson, et~al.]{phuong2024evaluating}
M~Phuong, M~Aitchison, E~Catt, S~Cogan, A~Kaskasoli, V~Krakovna, D~Lindner, M~Rahtz, Y~Assael, S~Hodkinson, et~al.
\newblock Evaluating frontier models for dangerous capabilities. arxiv.
\newblock \emph{arXiv preprint arXiv:2403.13793}, 2024.

\bibitem[Qi et~al.(2024)Qi, Wei, Carlini, Huang, Xie, He, Jagielski, Nasr, Mittal, and Henderson]{qi2024evaluating}
Xiangyu Qi, Boyi Wei, Nicholas Carlini, Yangsibo Huang, Tinghao Xie, Luxi He, Matthew Jagielski, Milad Nasr, Prateek Mittal, and Peter Henderson.
\newblock On evaluating the durability of safeguards for open-weight llms.
\newblock \emph{arXiv preprint arXiv:2412.07097}, 2024.

\bibitem[Shao et~al.(2024)Shao, Jancheska, Udeshi, Dolan-Gavitt, Milner, Chen, Yin, Garg, Krishnamurthy, Khorrami, et~al.]{shao2024nyu}
Minghao Shao, Sofija Jancheska, Meet Udeshi, Brendan Dolan-Gavitt, Kimberly Milner, Boyuan Chen, Max Yin, Siddharth Garg, Prashanth Krishnamurthy, Farshad Khorrami, et~al.
\newblock Nyu ctf bench: A scalable open-source benchmark dataset for evaluating llms in offensive security.
\newblock \emph{Advances in Neural Information Processing Systems}, 37:\penalty0 57472--57498, 2024.

\bibitem[Song \& Alves-Foss(2015)Song and Alves-Foss]{song2015darpa}
Jia Song and Jim Alves-Foss.
\newblock The darpa cyber grand challenge: A competitor's perspective.
\newblock \emph{IEEE Security \& Privacy}, 13\penalty0 (6):\penalty0 72--76, 2015.

\bibitem[Team et~al.(2025)Team, Bai, Bao, Chen, Chen, Chen, Chen, Chen, Chen, Chen, et~al.]{team2025kimi}
Kimi Team, Yifan Bai, Yiping Bao, Guanduo Chen, Jiahao Chen, Ningxin Chen, Ruijue Chen, Yanru Chen, Yuankun Chen, Yutian Chen, et~al.
\newblock Kimi k2: Open agentic intelligence.
\newblock \emph{arXiv preprint arXiv:2507.20534}, 2025.

\bibitem[Tihanyi et~al.(2024)Tihanyi, Ferrag, Jain, Bisztray, and Debbah]{tihanyi2024cybermetric}
Norbert Tihanyi, Mohamed~Amine Ferrag, Ridhi Jain, Tamas Bisztray, and Merouane Debbah.
\newblock Cybermetric: a benchmark dataset based on retrieval-augmented generation for evaluating llms in cybersecurity knowledge.
\newblock In \emph{2024 IEEE International Conference on Cyber Security and Resilience (CSR)}, pp.\  296--302. IEEE, 2024.

\bibitem[Wan et~al.(2024)Wan, Nikolaidis, Song, Molnar, Crnkovich, Grace, Bhatt, Chennabasappa, Whitman, Ding, et~al.]{wan2024cyberseceval}
Shengye Wan, Cyrus Nikolaidis, Daniel Song, David Molnar, James Crnkovich, Jayson Grace, Manish Bhatt, Sahana Chennabasappa, Spencer Whitman, Stephanie Ding, et~al.
\newblock Cyberseceval 3: Advancing the evaluation of cybersecurity risks and capabilities in large language models.
\newblock \emph{arXiv preprint arXiv:2408.01605}, 2024.

\bibitem[Wang et~al.(2025{\natexlab{a}})Wang, Liu, and Xiao]{wang2025cve}
Peiran Wang, Xiaogeng Liu, and Chaowei Xiao.
\newblock Cve-bench: Benchmarking llm-based software engineering agent’s ability to repair real-world cve vulnerabilities.
\newblock In \emph{Proceedings of the 2025 Conference of the Nations of the Americas Chapter of the Association for Computational Linguistics: Human Language Technologies (Volume 1: Long Papers)}, pp.\  4207--4224, 2025{\natexlab{a}}.

\bibitem[Wang et~al.(2025{\natexlab{b}})Wang, Shi, He, Cai, Zhang, and Song]{wang2025cybergym}
Zhun Wang, Tianneng Shi, Jingxuan He, Matthew Cai, Jialin Zhang, and Dawn Song.
\newblock Cybergym: Evaluating ai agents' cybersecurity capabilities with real-world vulnerabilities at scale.
\newblock \emph{arXiv preprint arXiv:2506.02548}, 2025{\natexlab{b}}.

\bibitem[Wei et~al.(2024)Wei, Wang, Liu, Ding, and Zhang]{wei2024magicoder}
Yuxiang Wei, Zhe Wang, Jiawei Liu, Yifeng Ding, and Lingming Zhang.
\newblock Magicoder: Empowering code generation with oss-instruct.
\newblock In \emph{International Conference on Machine Learning}, pp.\  52632--52657. PMLR, 2024.

\bibitem[Wei et~al.(2025)Wei, Duchenne, Copet, Carbonneaux, Zhang, Fried, Synnaeve, Singh, and Wang]{wei2025swe}
Yuxiang Wei, Olivier Duchenne, Jade Copet, Quentin Carbonneaux, Lingming Zhang, Daniel Fried, Gabriel Synnaeve, Rishabh Singh, and Sida~I Wang.
\newblock Swe-rl: Advancing llm reasoning via reinforcement learning on open software evolution.
\newblock \emph{arXiv preprint arXiv:2502.18449}, 2025.

\bibitem[{xAI}(2025)]{xai2025_rmfdraft}
{xAI}.
\newblock {xAI Risk Management Framework (Draft)}.
\newblock Technical report, xAI, February 2025.
\newblock Draft version — accessed 3 July 2025.

\bibitem[Xie et~al.(2025)Xie, Li, Gao, Du, Lam, Zou, and Chen]{xie2025swe}
Chengxing Xie, Bowen Li, Chang Gao, He~Du, Wai Lam, Difan Zou, and Kai Chen.
\newblock Swe-fixer: Training open-source llms for effective and efficient github issue resolution.
\newblock \emph{arXiv preprint arXiv:2501.05040}, 2025.

\bibitem[Yang et~al.(2025{\natexlab{a}})Yang, Li, Yang, Zhang, Hui, Zheng, Yu, Gao, Huang, Lv, et~al.]{yang2025qwen3}
An~Yang, Anfeng Li, Baosong Yang, Beichen Zhang, Binyuan Hui, Bo~Zheng, Bowen Yu, Chang Gao, Chengen Huang, Chenxu Lv, et~al.
\newblock Qwen3 technical report.
\newblock \emph{arXiv preprint arXiv:2505.09388}, 2025{\natexlab{a}}.

\bibitem[Yang et~al.(2023)Yang, Prabhakar, Narasimhan, and Yao]{yang2023intercode}
John Yang, Akshara Prabhakar, Karthik Narasimhan, and Shunyu Yao.
\newblock Intercode: Standardizing and benchmarking interactive coding with execution feedback.
\newblock \emph{Advances in Neural Information Processing Systems}, 36:\penalty0 23826--23854, 2023.

\bibitem[Yang et~al.(2024{\natexlab{a}})Yang, Jimenez, Wettig, Lieret, Yao, Narasimhan, and Press]{yang2024swe}
John Yang, Carlos~E Jimenez, Alexander Wettig, Kilian Lieret, Shunyu Yao, Karthik Narasimhan, and Ofir Press.
\newblock Swe-agent: Agent-computer interfaces enable automated software engineering.
\newblock \emph{Advances in Neural Information Processing Systems}, 37:\penalty0 50528--50652, 2024{\natexlab{a}}.

\bibitem[Yang et~al.(2025{\natexlab{b}})Yang, Leret, Jimenez, Wettig, Khandpur, Zhang, Hui, Press, Schmidt, and Yang]{yang2025swe}
John Yang, Kilian Leret, Carlos~E Jimenez, Alexander Wettig, Kabir Khandpur, Yanzhe Zhang, Binyuan Hui, Ofir Press, Ludwig Schmidt, and Diyi Yang.
\newblock Swe-smith: Scaling data for software engineering agents.
\newblock \emph{arXiv preprint arXiv:2504.21798}, 2025{\natexlab{b}}.

\bibitem[Yang et~al.(2024{\natexlab{b}})Yang, Nie, Wang, Tang, Guo, Li, and Song]{yang2024seccodeplt}
Yu~Yang, Yuzhou Nie, Zhun Wang, Yuheng Tang, Wenbo Guo, Bo~Li, and Dawn Song.
\newblock Seccodeplt: A unified platform for evaluating the security of code genai.
\newblock \emph{arXiv preprint arXiv:2410.11096}, 2024{\natexlab{b}}.

\bibitem[Zhang et~al.(2025{\natexlab{a}})Zhang, Ji, Menders, Dulepet, Qin, Wang, Wu, Liao, Li, Hu, et~al.]{zhang2025bountybench}
Andy~K Zhang, Joey Ji, Celeste Menders, Riya Dulepet, Thomas Qin, Ron~Y Wang, Junrong Wu, Kyleen Liao, Jiliang Li, Jinghan Hu, et~al.
\newblock Bountybench: Dollar impact of ai agent attackers and defenders on real-world cybersecurity systems.
\newblock \emph{arXiv preprint arXiv:2505.15216}, 2025{\natexlab{a}}.

\bibitem[Zhang et~al.(2025{\natexlab{b}})Zhang, Perry, Dulepet, Ji, Menders, Lin, Jones, Hussein, Liu, Jasper, Peetathawatchai, Glenn, Sivashankar, Zamoshchin, Glikbarg, Askaryar, Yang, Zhang, Alluri, Tran, Sangpisit, Oseleononmen, Boneh, Ho, and Liang]{zhang2025cybench}
Andy~K Zhang, Neil Perry, Riya Dulepet, Joey Ji, Celeste Menders, Justin~W Lin, Eliot Jones, Gashon Hussein, Samantha Liu, Donovan~Julian Jasper, Pura Peetathawatchai, Ari Glenn, Vikram Sivashankar, Daniel Zamoshchin, Leo Glikbarg, Derek Askaryar, Haoxiang Yang, Aolin Zhang, Rishi Alluri, Nathan Tran, Rinnara Sangpisit, Kenny~O Oseleononmen, Dan Boneh, Daniel~E. Ho, and Percy Liang.
\newblock Cybench: A framework for evaluating cybersecurity capabilities and risks of language models.
\newblock In \emph{The Thirteenth International Conference on Learning Representations}, 2025{\natexlab{b}}.
\newblock URL \url{https://openreview.net/forum?id=tc90LV0yRL}.

\bibitem[Zhu et~al.(2025)Zhu, Kellermann, Bowman, Li, Gupta, Danda, Fang, Jensen, Ihli, Benn, et~al.]{zhucve}
Yuxuan Zhu, Antony Kellermann, Dylan Bowman, Philip Li, Akul Gupta, Adarsh Danda, Richard Fang, Conner Jensen, Eric Ihli, Jason Benn, et~al.
\newblock Cve-bench: A benchmark for ai agents’ ability to exploit real-world web application vulnerabilities.
\newblock In \emph{Forty-second International Conference on Machine Learning}, 2025.

\bibitem[Zhuo et~al.(2024)Zhuo, Zebaze, Suppattarachai, von Werra, de~Vries, Liu, and Muennighoff]{zhuo2024astraios}
Terry~Yue Zhuo, Armel Zebaze, Nitchakarn Suppattarachai, Leandro von Werra, Harm de~Vries, Qian Liu, and Niklas Muennighoff.
\newblock Astraios: Parameter-efficient instruction tuning code large language models.
\newblock \emph{arXiv preprint arXiv:2401.00788}, 2024.

\bibitem[Zhuo et~al.(2025)Zhuo, Wang, Ding, Kumar, and Wang]{zhuo2025cyber}
Terry~Yue Zhuo, Dingmin Wang, Hantian Ding, Varun Kumar, and Zijian Wang.
\newblock Cyber-zero: Training cybersecurity agents without runtime.
\newblock \emph{arXiv preprint}, 2025.

\end{thebibliography}

\restoreTOC

\clearpage
\onecolumn  %
\appendix
\part*{Appendix} %

\begingroup
 \let\clearpage\relax %
 \tableofcontents
\endgroup

\clearpage

\section{Statistics}
We provide a summary of the important statistics mentioned in the paper.
\begin{table}[!h]
\centering
\caption{Summary of data statistics.}
\resizebox{0.7\linewidth}{!}{
\begin{tabular}{p{0.55\linewidth} r}
\toprule
\textbf{Item Description} & \textbf{Count} \\
\midrule
\multicolumn{2}{c}{\textit{\dojo Challenges}} \\
\midrule
Number of available CTF challenges & 658 \\
Number of challenges with stable and reproducible environments, as confirmed by the original authors & 650 \\
\midrule
\multicolumn{2}{c}{\textit{Writeups for CTF Challenges}} \\
\midrule
Total number of writeups collected from the CTFtime website & 8,361 \\
Writeups successfully matched to \dojo{} challenges using competition and task metadata & 252 \\
\dojo{} challenges for which at least one corresponding writeup is available & 150 \\
\midrule
\multicolumn{2}{c}{\textit{Successful Agent Samples}} \\
\midrule
Raw agent trajectories collected before cleaning or filtering & 1,006 \\
Unique trajectories remaining after removing duplicates and limiting the maximum number per challenge & 486 \\
\dojo{} challenges that include at least one valid and successful trajectory & 274 \\
\bottomrule
\end{tabular}}
\label{tab:stats}
\end{table}

\newpage

\section{Data Analysis}
\label{app:analysis}
\begin{figure*}[!h]
    \centering
    \includegraphics[width=0.9\linewidth]{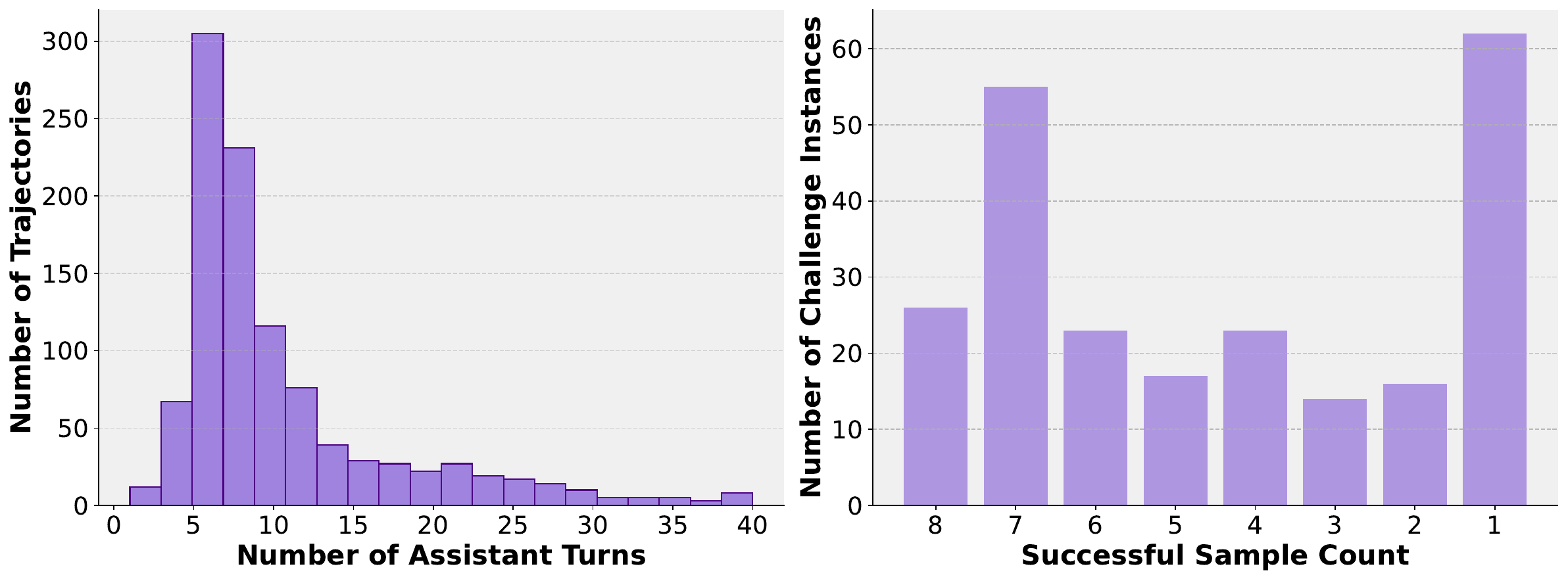}
    \caption{
    Number of turns in each successful trajectory (left) and number of successful trajectories for each challenge instance (right).}
    \label{fig:count_stats}
\end{figure*}

\autoref{fig:count_stats} presents two key statistics of the collected data. The left panel visualizes the number of assistant turns per trajectory. The majority of trajectories fall between 5 to 15 turns, with a heavy tail extending to 40 turns. This skew indicates that while many tasks can be solved efficiently, a substantial portion demands prolonged, iterative explorations, highlighting the complex nature of real-world CTF problems. The right panel plots
the number of successful trajectories obtained for each challenge, revealing that many challenges are solved only once within the total 12 rollouts, indicating that successful trajectories for certain instances are difficult to collect.

\section{Main Results}

\section{More Ablation Studies}
\label{app:ablation}

\begin{wraptable}{r}{0.35\linewidth}
\vspace{-10pt}
\centering
\resizebox{\linewidth}{!}{
\begin{tabular}{lccc}
\toprule
\textbf{Category} & \textbf{Qwen} & \cellcolor{cyan!20}\textbf{Both} & \textbf{DeepSeek} \\
\midrule
\textbf{Crypto}    & 31 & \cellcolor{cyan!20}\textbf{84} & 26  \\
\textbf{Forensics} &  1 & \cellcolor{cyan!20}\textbf{13} &  3 \\
\textbf{Pwn}       &  2 & \cellcolor{cyan!20}\textbf{15} &  3 \\
\textbf{Rev}       &  6 & \cellcolor{cyan!20}\textbf{37} &  9 \\
\textbf{Web}       &  0 & \cellcolor{cyan!20}\textbf{6}  &  2 \\
\textbf{Misc}      &  4 & \cellcolor{cyan!20}\textbf{26} &  6  \\
\bottomrule
\end{tabular}}
\caption{Solved challenge counts.}
\vspace{-10pt}
\label{tab:categories}
\end{wraptable}

\paragraph{Setup}
To assess the benefit of using multiple teacher models during trajectory collection, we compare the individual and combined contributions of Qwen3-Coder and DeepSeek-V3-0324. We first analyze how many unique challenges each model solves and their category-level overlaps. Then, we fine-tune Qwen3 models of sizes 8B, 14B, and 32B on three trajectory subsets: (1) Qwen3-Coder only, (2) DeepSeek-V3-0324 only, and (3) both combined. We report average Pass@1 across benchmarks to evaluate downstream agent performance. Decoding parameters and training setup match those in our main experiments.

\paragraph{Analysis}
In \autoref{tab:categories}, Qwen3-Coder and DeepSeek-V3-0324 demonstrate complementary strengths. For example, in Crypto tasks, the models share 84 solves, but Qwen3-Coder uniquely solves 31 while DeepSeek-V3-0324 adds another 26. Similar patterns emerge across other categories, with notable non-overlapping contributions in Reverse Engineering, Misc, and Forensics. Combining both models increases total coverage to 274 unique challenges, exceeding either model alone. This diversity translates into measurable downstream gains. 
\begin{wraptable}{r}{0.4\linewidth}
  \centering
  \caption{Pass@1 performance when varying teacher models.}
  \label{tab:teacher_ablation}
  \resizebox{\linewidth}{!}{
  \begin{tabular}{l|ccc}
    \toprule
    \textbf{Teacher Model} & \textbf{8B} & \textbf{14B} & \textbf{32B} \\
    \midrule
    Qwen3-Coder      & 17.3 & 23.8 & 29.4 \\
    DeepSeek-V3-0324          & 17.6 & 24.8 & 31.3 \\ \midrule
    Combined       & \textbf{18.9} & \textbf{25.7} & \textbf{31.9} \\
    \bottomrule
  \end{tabular}
  }
\end{wraptable}

\autoref{tab:teacher_ablation} reveals that training on combined trajectories improves Pass@1 performance across all model sizes. For example, the 32B model trained on combined data achieves 31.9\%, outperforming both the Qwen3-Coder-only (29.4\%) and DeepSeek-only (31.3\%) variants. Similarly, the 8B and 14B models also benefit from the combined setting. These results confirm that teacher diversity enriches training data and yields more capable cybersecurity agents.

\section{More Related Work}
\paragraph{Training LLM Agents to Code}
Previous training paradigms for software engineering have largely emphasized general-purpose coding capabilities~\citep{li2023starcoder,lozhkov2024starcoder,muennighoffoctopack,zhuo2024astraios,wei2024magicoder}. While scaffolded approaches using proprietary models achieve strong results on real-world software engineering (SE) tasks, open-source models continue to lag behind, prompting a shift toward domain-specific training strategies. Several recent efforts exemplify this trend. Lingma SWE-GPT~\citep{ma2024lingma} introduces 7B and 72B models trained with a process-oriented development methodology. SWE-Gym~\citep{pan2024training} offers the first open training environment for SE agents, yielding notable gains on SWE-bench~\citep{jimenezswe}. More recent work includes SWE-smith~\citep{yang2025swe}, which automatically scales training data for SE, and SWE-RL~\citep{wei2025swe}, which applies reinforcement learning~\citep{grattafiori2024llama} to repair programs with reasoning.
While these methods advance software engineering capabilities via execution-based environments, they do not address the distinct demands of cybersecurity~\citep{zhuo2025cyber}. Our work fills this gap by introducing the first execution environment specifically tailored for security tasks, where traditional code-centric training fails to transfer effectively.

\section{\dojo CTF Challenges}

\begin{longtable}{llllll}
\toprule
Competition & Challenge & Category & Qwen & DeepSeek \\
\midrule
\endfirsthead
\multicolumn{5}{c}%
{\tablename\ \thetable{} -- \textit{Continued from previous page}} \\
\toprule
Competition & Challenge & Category & Qwen & DeepSeek \\
\midrule
\endhead
\midrule
\multicolumn{5}{r}{\textit{Continued on next page}} \\
\endfoot
\bottomrule
\endlastfoot
\nopagebreak[4]
\multirow{3}{*}{0CTF - 2017}
 & babyheap & Pwn & \cmark & \xmark \\
\nopagebreak[4]

 & diethard & Pwn & \cmark & \xmark \\
\nopagebreak[4]

 & easiestprintf & Pwn & \xmark & \xmark \\
\midrule
\nopagebreak[4]
\multirow{6}{*}{0CTF - 2018}
 & babyheap2018 & Pwn & \xmark & \cmark \\
\nopagebreak[4]

 & blackhole & Pwn & \xmark & \cmark \\
\nopagebreak[4]

 & freenote2018 & Pwn & \xmark & \xmark \\
\nopagebreak[4]

 & heapstorm & Pwn & \xmark & \xmark \\
\nopagebreak[4]

 & subtraction & Misc & \cmark & \xmark \\
\nopagebreak[4]

 & zerofs & Pwn & \xmark & \xmark \\
\midrule
\nopagebreak[4]
\multirow{10}{*}{0CTF - 2019}
 & babyaegis & Pwn & \xmark & \cmark \\
\nopagebreak[4]

 & babyheap & Pwn & \cmark & \cmark \\
\nopagebreak[4]

 & babyrsa & Crypto & \cmark & \xmark \\
\nopagebreak[4]

 & babysandbox & Pwn & \xmark & \xmark \\
\nopagebreak[4]

 & elements & Rev & \cmark & \xmark \\
\nopagebreak[4]

 & flropyd & Pwn & \xmark & \xmark \\
\nopagebreak[4]

 & plang & Pwn & \xmark & \xmark \\
\nopagebreak[4]

 & sanitize & Misc & \cmark & \xmark \\
\nopagebreak[4]

 & scanner & Pwn & \xmark & \xmark \\
\nopagebreak[4]

 & zerotask & Pwn & \xmark & \xmark \\
\midrule
\nopagebreak[4]
\multirow{5}{*}{0CTF Quals - 2021}
 & cloudpass & Crypto & \cmark & \xmark \\
\nopagebreak[4]

 & future & Rev & \cmark & \xmark \\
\nopagebreak[4]

 & listbook & Pwn & \cmark & \cmark \\
\nopagebreak[4]

 & vp & Rev & \cmark & \xmark \\
\nopagebreak[4]

 & zer0lfsr & Crypto & \cmark & \xmark \\
\pagebreak
\nopagebreak[4]
\multirow{17}{*}{0xCTF - 4141}
 & client & Rev & \cmark & \xmark \\
\nopagebreak[4]

 & eazyrsa & Crypto & \cmark & \xmark \\
\nopagebreak[4]

 & external & Pwn & \cmark & \cmark \\
\nopagebreak[4]

 & factorize & Crypto & \cmark & \xmark \\
\nopagebreak[4]

 & filereader & Misc & \cmark & \xmark \\
\nopagebreak[4]

 & hash & Rev & \cmark & \xmark \\
\nopagebreak[4]

 & moving-signals & Pwn & \cmark & \cmark \\
\nopagebreak[4]

 & pyjail & Misc & \cmark & \xmark \\
\nopagebreak[4]

 & ret-of-the-rops & Pwn & \xmark & \xmark \\
\nopagebreak[4]

 & shjail & Misc & \xmark & \xmark \\
\nopagebreak[4]

 & soul & Crypto & \cmark & \xmark \\
\nopagebreak[4]

 & staple-aes & Crypto & \xmark & \xmark \\
\nopagebreak[4]

 & the-pwn-inn & Pwn & \xmark & \xmark \\
\nopagebreak[4]

 & wallet & Crypto & \xmark & \xmark \\
\nopagebreak[4]

 & ware & Rev & \xmark & \xmark \\
\nopagebreak[4]

 & wrongdownload & Rev & \xmark & \xmark \\
\nopagebreak[4]

 & x-and-or & Rev & \xmark & \xmark \\
\midrule
\nopagebreak[4]
\multirow{7}{*}{29c3CTF - 2012}
 & findthekey & Rev & \cmark & \xmark \\
\nopagebreak[4]

 & maya & Rev & \xmark & \cmark \\
\nopagebreak[4]

 & memcached & Pwn & \cmark & \cmark \\
\nopagebreak[4]

 & minesweeper & Pwn & \cmark & \cmark \\
\nopagebreak[4]

 & proxy & Pwn & \xmark & \xmark \\
\nopagebreak[4]

 & ru1337 & Pwn & \xmark & \xmark \\
\nopagebreak[4]

 & updateserver & Pwn & \xmark & \xmark \\
\midrule
\nopagebreak[4]
\multirow{12}{*}{AccessdeniedCTF - 2022}
 & babyc & Misc & \xmark & \cmark \\
\nopagebreak[4]

 & binary & Rev & \xmark & \cmark \\
\nopagebreak[4]

 & ecc & Crypto & \cmark & \xmark \\
\nopagebreak[4]

 & enormous & Rev & \xmark & \cmark \\
\nopagebreak[4]

 & llvm & Rev & \xmark & \xmark \\
\nopagebreak[4]

 & merklegoodman & Crypto & \cmark & \xmark \\
\nopagebreak[4]

 & mitm2 & Crypto & \cmark & \xmark \\
\nopagebreak[4]

 & ret2system & Pwn & \cmark & \cmark \\
\nopagebreak[4]

 & rsa1 & Crypto & \xmark & \xmark \\
\nopagebreak[4]

 & rsa2 & Crypto & \xmark & \xmark \\
\nopagebreak[4]

 & rsa3 & Crypto & \xmark & \xmark \\
\nopagebreak[4]

 & smallkey & Crypto & \xmark & \xmark \\
\pagebreak
\nopagebreak[4]
\multirow{25}{*}{AngstromCTF - 2016}
 & amoebananas & Web & \xmark & \cmark \\
\nopagebreak[4]

 & artifact & Crypto & \cmark & \xmark \\
\nopagebreak[4]

 & asmtracing & Rev & \xmark & \cmark \\
\nopagebreak[4]

 & casino & Crypto & \cmark & \xmark \\
\nopagebreak[4]

 & cipher & Rev & \xmark & \cmark \\
\nopagebreak[4]

 & ciphertwo & Rev & \xmark & \xmark \\
\nopagebreak[4]

 & client & Web & \xmark & \cmark \\
\nopagebreak[4]

 & drag & Misc & \xmark & \cmark \\
\nopagebreak[4]

 & endian & Pwn & \cmark & \cmark \\
\nopagebreak[4]

 & fender & Forensics & \cmark & \xmark \\
\nopagebreak[4]

 & flaglock & Misc & \xmark & \cmark \\
\nopagebreak[4]

 & formatone & Pwn & \cmark & \cmark \\
\nopagebreak[4]

 & hamlet & Crypto & \cmark & \xmark \\
\nopagebreak[4]

 & headsup & Forensics & \xmark & \cmark \\
\nopagebreak[4]

 & helpcenter & Crypto & \xmark & \xmark \\
\nopagebreak[4]

 & hex & Crypto & \xmark & \xmark \\
\nopagebreak[4]

 & imageencryptor & Rev & \xmark & \xmark \\
\nopagebreak[4]

 & javabest & Rev & \xmark & \xmark \\
\nopagebreak[4]

 & metasploit & Forensics & \xmark & \xmark \\
\nopagebreak[4]

 & music & Forensics & \xmark & \xmark \\
\nopagebreak[4]

 & oops & Forensics & \xmark & \xmark \\
\nopagebreak[4]

 & recovery & Forensics & \xmark & \xmark \\
\nopagebreak[4]

 & rsa & Crypto & \xmark & \xmark \\
\nopagebreak[4]

 & spqr & Crypto & \xmark & \xmark \\
\nopagebreak[4]

 & yankovic & Forensics & \xmark & \xmark \\
\midrule
\nopagebreak[4]
\multirow{6}{*}{AngstromCTF - 2017}
 & begin & Crypto & \cmark & \xmark \\
\nopagebreak[4]

 & casino & Crypto & \cmark & \xmark \\
\nopagebreak[4]

 & knockknock & Crypto & \cmark & \xmark \\
\nopagebreak[4]

 & obligatory & Web & \cmark & \cmark \\
\nopagebreak[4]

 & royalcasino & Crypto & \xmark & \xmark \\
\nopagebreak[4]

 & substitutioncipher & Crypto & \xmark & \xmark \\
\pagebreak
\nopagebreak[4]
\multirow{13}{*}{AngstromCTF - 2018}
 & accumulator & Pwn & \cmark & \cmark \\
\nopagebreak[4]

 & backtobasics & Crypto & \cmark & \xmark \\
\nopagebreak[4]

 & bankroppery & Pwn & \cmark & \cmark \\
\nopagebreak[4]

 & introtorsa & Crypto & \cmark & \xmark \\
\nopagebreak[4]

 & productkey & Rev & \xmark & \cmark \\
\nopagebreak[4]

 & rev1 & Rev & \xmark & \cmark \\
\nopagebreak[4]

 & rev2 & Rev & \xmark & \xmark \\
\nopagebreak[4]

 & rev3 & Rev & \xmark & \xmark \\
\nopagebreak[4]

 & waldo2 & Misc & \xmark & \cmark \\
\nopagebreak[4]

 & warmup & Misc & \xmark & \cmark \\
\nopagebreak[4]

 & washington & Rev & \xmark & \xmark \\
\nopagebreak[4]

 & weirdmessage & Misc & \xmark & \xmark \\
\nopagebreak[4]

 & xor & Crypto & \cmark & \xmark \\
\midrule
\nopagebreak[4]
\multirow{11}{*}{AngstromCTF - 2019}
 & blankpaper & Misc & \xmark & \cmark \\
\nopagebreak[4]

 & chainofrope & Pwn & \cmark & \cmark \\
\nopagebreak[4]

 & highqualitychecks & Rev & \xmark & \cmark \\
\nopagebreak[4]

 & icthyo & Rev & \xmark & \cmark \\
\nopagebreak[4]

 & like & Rev & \xmark & \xmark \\
\nopagebreak[4]

 & lithp & Misc & \cmark & \cmark \\
\nopagebreak[4]

 & onebite & Rev & \xmark & \xmark \\
\nopagebreak[4]

 & overmybrain & Pwn & \cmark & \cmark \\
\nopagebreak[4]

 & paperbin & Misc & \xmark & \xmark \\
\nopagebreak[4]

 & reallysecurealgorithm & Crypto & \cmark & \xmark \\
\nopagebreak[4]

 & runes & Crypto & \cmark & \xmark \\
\midrule
\nopagebreak[4]
\multirow{10}{*}{AngstromCTF - 2022}
 & amongus & Misc & \cmark & \cmark \\
\nopagebreak[4]

 & caesaranddesister & Crypto & \cmark & \xmark \\
\nopagebreak[4]

 & dyn & Rev & \cmark & \cmark \\
\nopagebreak[4]

 & numbergame & Rev & \cmark & \cmark \\
\nopagebreak[4]

 & randomlysampledalgorithm & Crypto & \cmark & \xmark \\
\nopagebreak[4]

 & reallyobnoxiousproblem & Pwn & \cmark & \cmark \\
\nopagebreak[4]

 & shark1 & Misc & \cmark & \cmark \\
\nopagebreak[4]

 & uninspired & Rev & \xmark & \xmark \\
\nopagebreak[4]

 & wah & Pwn & \cmark & \cmark \\
\nopagebreak[4]

 & whatsmyname & Pwn & \xmark & \xmark \\
\pagebreak
\nopagebreak[4]
\multirow{15}{*}{AngstromCTF - 2024}
 & awman & Crypto & \cmark & \xmark \\
\nopagebreak[4]

 & bap & Pwn & \xmark & \xmark \\
\nopagebreak[4]

 & exam & Pwn & \xmark & \xmark \\
\nopagebreak[4]

 & heapify & Pwn & \xmark & \xmark \\
\nopagebreak[4]

 & layers & Misc & \cmark & \cmark \\
\nopagebreak[4]

 & leftright & Pwn & \xmark & \xmark \\
\nopagebreak[4]

 & og & Pwn & \xmark & \xmark \\
\nopagebreak[4]

 & philosophy & Crypto & \cmark & \xmark \\
\nopagebreak[4]

 & presidential & Pwn & \xmark & \xmark \\
\nopagebreak[4]

 & simonsays & Crypto & \cmark & \xmark \\
\nopagebreak[4]

 & snowman & Misc & \cmark & \cmark \\
\nopagebreak[4]

 & stacksort & Pwn & \xmark & \xmark \\
\nopagebreak[4]

 & themectl & Pwn & \xmark & \xmark \\
\nopagebreak[4]

 & tss1 & Crypto & \xmark & \xmark \\
\nopagebreak[4]

 & tss2 & Crypto & \xmark & \xmark \\
\midrule
\nopagebreak[4]
\multirow{9}{*}{AsisCTF - 2013}
 & dice & Rev & \cmark & \cmark \\
\nopagebreak[4]

 & encoding & Crypto & \cmark & \xmark \\
\nopagebreak[4]

 & inaccessible & Forensics & \xmark & \cmark \\
\nopagebreak[4]

 & licensekey & Rev & \cmark & \cmark \\
\nopagebreak[4]

 & memdump & Forensics & \xmark & \cmark \\
\nopagebreak[4]

 & pcaps & Crypto & \cmark & \xmark \\
\nopagebreak[4]

 & rsang & Crypto & \cmark & \xmark \\
\nopagebreak[4]

 & serialnumber & Rev & \xmark & \xmark \\
\nopagebreak[4]

 & simpleofficer & Crypto & \xmark & \xmark \\
\midrule
\nopagebreak[4]
\multirow{2}{*}{AsisCTF - 2014}
 & blocks & Forensics & \cmark & \cmark \\
\nopagebreak[4]

 & randomimage & Crypto & \cmark & \xmark \\
\midrule
\nopagebreak[4]
\multirow{8}{*}{BackdoorCTF - 2019}
 & babyheapbackdoorctf & Pwn & \xmark & \xmark \\
\nopagebreak[4]

 & babytcache & Pwn & \xmark & \xmark \\
\nopagebreak[4]

 & echo & Pwn & \xmark & \xmark \\
\nopagebreak[4]

 & forgot & Pwn & \xmark & \xmark \\
\nopagebreak[4]

 & matrix & Pwn & \xmark & \xmark \\
\nopagebreak[4]

 & miscpwn & Pwn & \xmark & \xmark \\
\nopagebreak[4]

 & rsanne & Crypto & \cmark & \xmark \\
\nopagebreak[4]

 & team & Pwn & \xmark & \xmark \\
\pagebreak
\nopagebreak[4]
\multirow{11}{*}{ByuCTF - 2022}
 & ballgame & Crypto & \cmark & \xmark \\
\nopagebreak[4]

 & basicrev & Rev & \cmark & \cmark \\
\nopagebreak[4]

 & blue & Forensics & \cmark & \cmark \\
\nopagebreak[4]

 & chicken & Rev & \cmark & \cmark \\
\nopagebreak[4]

 & funfact & Rev & \xmark & \xmark \\
\nopagebreak[4]

 & murdermystery & Misc & \cmark & \cmark \\
\nopagebreak[4]

 & qool & Forensics & \cmark & \cmark \\
\nopagebreak[4]

 & shift & Crypto & \xmark & \cmark \\
\nopagebreak[4]

 & stickykey & Forensics & \xmark & \xmark \\
\nopagebreak[4]

 & truth & Crypto & \xmark & \cmark \\
\nopagebreak[4]

 & xqr & Crypto & \xmark & \xmark \\
\midrule
\nopagebreak[4]
\multirow{12}{*}{ByuCTF - 2023}
 & crcconfusion & Forensics & \cmark & \cmark \\
\nopagebreak[4]

 & hexadecalingo & Misc & \cmark & \cmark \\
\nopagebreak[4]

 & misc006-1 & Misc & \cmark & \cmark \\
\nopagebreak[4]

 & misc006-2 & Misc & \xmark & \xmark \\
\nopagebreak[4]

 & poem & Crypto & \xmark & \cmark \\
\nopagebreak[4]

 & pwn2038 & Pwn & \xmark & \xmark \\
\nopagebreak[4]

 & rsa1 & Crypto & \xmark & \cmark \\
\nopagebreak[4]

 & rsa2 & Crypto & \xmark & \cmark \\
\nopagebreak[4]

 & rsa3 & Crypto & \xmark & \xmark \\
\nopagebreak[4]

 & rsa4 & Crypto & \xmark & \xmark \\
\nopagebreak[4]

 & rsa5 & Crypto & \xmark & \xmark \\
\nopagebreak[4]

 & xkcd2637 & Misc & \xmark & \xmark \\
\midrule
\nopagebreak[4]
\multirow{11}{*}{ByuCTF - 2024}
 & aresa & Crypto & \xmark & \cmark \\
\nopagebreak[4]

 & domath & Crypto & \xmark & \cmark \\
\nopagebreak[4]

 & giveup & Crypto & \xmark & \cmark \\
\nopagebreak[4]

 & gotmail & Misc & \cmark & \cmark \\
\nopagebreak[4]

 & meetgreg & Misc & \cmark & \cmark \\
\nopagebreak[4]

 & multiplied & Crypto & \xmark & \xmark \\
\nopagebreak[4]

 & petrolhead & Misc & \xmark & \xmark \\
\nopagebreak[4]

 & typosquatting & Misc & \xmark & \xmark \\
\nopagebreak[4]

 & vacationboats & Misc & \xmark & \xmark \\
\nopagebreak[4]

 & wateryoudoing & Misc & \xmark & \xmark \\
\nopagebreak[4]

 & worstchallenge & Forensics & \cmark & \cmark \\
\midrule
\nopagebreak[4]
\multirow{6}{*}{CactusconCTF - 2025}
 & clueless & Misc & \cmark & \cmark \\
\nopagebreak[4]

 & frng & Misc & \cmark & \cmark \\
\nopagebreak[4]

 & numbersleuthv1 & Misc & \xmark & \xmark \\
\nopagebreak[4]

 & numbersleuthv2 & Misc & \xmark & \xmark \\
\nopagebreak[4]

 & numbersleuthv3 & Misc & \xmark & \xmark \\
\nopagebreak[4]

 & securerepititions & Misc & \xmark & \xmark \\
\pagebreak
\nopagebreak[4]
\multirow{6}{*}{CcscCTF - 2020}
 & basilisk64 & Crypto & \xmark & \cmark \\
\nopagebreak[4]

 & echoes & Misc & \cmark & \cmark \\
\nopagebreak[4]

 & guy & Pwn & \xmark & \xmark \\
\nopagebreak[4]

 & mouse & Crypto & \xmark & \cmark \\
\nopagebreak[4]

 & routes & Crypto & \xmark & \cmark \\
\nopagebreak[4]

 & spell & Pwn & \xmark & \xmark \\
\midrule
\nopagebreak[4]
\multirow{13}{*}{Codegate - 2011}
 & binary100 & Pwn & \xmark & \xmark \\
\nopagebreak[4]

 & binary200 & Pwn & \xmark & \xmark \\
\nopagebreak[4]

 & binary300 & Pwn & \xmark & \xmark \\
\nopagebreak[4]

 & binary400 & Pwn & \xmark & \xmark \\
\nopagebreak[4]

 & binary500 & Pwn & \xmark & \xmark \\
\nopagebreak[4]

 & crypto200 & Crypto & \xmark & \cmark \\
\nopagebreak[4]

 & crypto300 & Crypto & \xmark & \cmark \\
\nopagebreak[4]

 & crypto400 & Crypto & \xmark & \cmark \\
\nopagebreak[4]

 & crypto500 & Crypto & \xmark & \xmark \\
\nopagebreak[4]

 & forensics200 & Forensics & \cmark & \cmark \\
\nopagebreak[4]

 & forensics300 & Forensics & \cmark & \cmark \\
\nopagebreak[4]

 & forensics400 & Forensics & \xmark & \xmark \\
\nopagebreak[4]

 & network100 & Web & \cmark & \cmark \\
\midrule
\nopagebreak[4]
\multirow{10}{*}{CodegateCTF - 2012}
 & bin100 & Pwn & \xmark & \xmark \\
\nopagebreak[4]

 & bin200 & Pwn & \xmark & \xmark \\
\nopagebreak[4]

 & bin300 & Pwn & \xmark & \xmark \\
\nopagebreak[4]

 & bin400 & Pwn & \xmark & \xmark \\
\nopagebreak[4]

 & bin500 & Pwn & \xmark & \xmark \\
\nopagebreak[4]

 & forensics100 & Forensics & \cmark & \cmark \\
\nopagebreak[4]

 & forensics200 & Forensics & \cmark & \cmark \\
\nopagebreak[4]

 & forensics300 & Forensics & \xmark & \xmark \\
\nopagebreak[4]

 & forensics400 & Misc & \cmark & \cmark \\
\nopagebreak[4]

 & vuln500 & Pwn & \xmark & \xmark \\
\midrule
CodegateCTF - 2013
 & vuln100 & Pwn & \xmark & \xmark \\
\midrule
\nopagebreak[4]
\multirow{9}{*}{Codegateprelims - 2014}
 & 4stone & Pwn & \xmark & \xmark \\
\nopagebreak[4]

 & angrydoraemon & Pwn & \xmark & \xmark \\
\nopagebreak[4]

 & automata & Rev & \cmark & \cmark \\
\nopagebreak[4]

 & chronological & Misc & \cmark & \cmark \\
\nopagebreak[4]

 & crackme & Rev & \cmark & \cmark \\
\nopagebreak[4]

 & dodosandbox & Pwn & \xmark & \xmark \\
\nopagebreak[4]

 & hypercat & Pwn & \xmark & \xmark \\
\nopagebreak[4]

 & minibomb & Pwn & \xmark & \xmark \\
\nopagebreak[4]

 & weirdsnus & Pwn & \xmark & \xmark \\
\pagebreak
\nopagebreak[4]
\multirow{11}{*}{CorCTF - 2021}
 & babyrand & Crypto & \xmark & \cmark \\
\nopagebreak[4]

 & babyrev & Rev & \cmark & \cmark \\
\nopagebreak[4]

 & bank & Crypto & \xmark & \cmark \\
\nopagebreak[4]

 & chainblock & Pwn & \xmark & \xmark \\
\nopagebreak[4]

 & chance & Crypto & \xmark & \cmark \\
\nopagebreak[4]

 & cshell & Pwn & \xmark & \xmark \\
\nopagebreak[4]

 & fibinary & Crypto & \xmark & \xmark \\
\nopagebreak[4]

 & fourninesix & Crypto & \xmark & \xmark \\
\nopagebreak[4]

 & friedrice & Crypto & \xmark & \xmark \\
\nopagebreak[4]

 & lcg & Crypto & \xmark & \xmark \\
\nopagebreak[4]

 & vmquack & Rev & \cmark & \cmark \\
\midrule
\nopagebreak[4]
\multirow{7}{*}{CorCTF - 2022}
 & babypad & Misc & \cmark & \cmark \\
\nopagebreak[4]

 & bogus & Rev & \cmark & \cmark \\
\nopagebreak[4]

 & edgelord & Rev & \cmark & \cmark \\
\nopagebreak[4]

 & exchanged & Crypto & \xmark & \cmark \\
\nopagebreak[4]

 & msfrob & Rev & \xmark & \xmark \\
\nopagebreak[4]

 & turbocrab & Rev & \xmark & \xmark \\
\nopagebreak[4]

 & vmquacksrevenge & Rev & \xmark & \xmark \\
\midrule
\nopagebreak[4]
\multirow{6}{*}{CryptoCTF - 2020}
 & amsterdam & Crypto & \xmark & \cmark \\
\nopagebreak[4]

 & complextohell & Crypto & \xmark & \cmark \\
\nopagebreak[4]

 & fatima & Crypto & \xmark & \cmark \\
\nopagebreak[4]

 & onelinecrypto & Crypto & \xmark & \xmark \\
\nopagebreak[4]

 & threeravens & Crypto & \xmark & \xmark \\
\nopagebreak[4]

 & trailingbits & Crypto & \xmark & \xmark \\
\midrule
\nopagebreak[4]
\multirow{14}{*}{CryptoCTF - 2021}
 & dorsa & Crypto & \xmark & \cmark \\
\nopagebreak[4]

 & ecchimera & Crypto & \xmark & \cmark \\
\nopagebreak[4]

 & elegant & Crypto & \xmark & \cmark \\
\nopagebreak[4]

 & farm & Crypto & \xmark & \xmark \\
\nopagebreak[4]

 & frozen & Crypto & \xmark & \xmark \\
\nopagebreak[4]

 & hamul & Crypto & \xmark & \xmark \\
\nopagebreak[4]

 & hypernormal & Crypto & \xmark & \xmark \\
\nopagebreak[4]

 & improved & Crypto & \xmark & \xmark \\
\nopagebreak[4]

 & lower & Crypto & \xmark & \xmark \\
\nopagebreak[4]

 & rima & Crypto & \xmark & \xmark \\
\nopagebreak[4]

 & tinyecc & Crypto & \xmark & \xmark \\
\nopagebreak[4]

 & triplet & Crypto & \xmark & \xmark \\
\nopagebreak[4]

 & trunc & Crypto & \xmark & \xmark \\
\nopagebreak[4]

 & wolf & Crypto & \xmark & \xmark \\
\pagebreak
\nopagebreak[4]
\multirow{6}{*}{CryptoverseCTF - 2022}
 & bigrabin & Crypto & \xmark & \cmark \\
\nopagebreak[4]

 & dlog & Crypto & \xmark & \cmark \\
\nopagebreak[4]

 & rsa2 & Crypto & \cmark & \cmark \\
\nopagebreak[4]

 & rsa3 & Crypto & \xmark & \xmark \\
\nopagebreak[4]

 & tale & Crypto & \xmark & \xmark \\
\nopagebreak[4]

 & worldcup & Rev & \cmark & \cmark \\
\midrule
\nopagebreak[4]
\multirow{11}{*}{CryptoverseCTF - 2023}
 & acceptance & Pwn & \xmark & \xmark \\
\nopagebreak[4]

 & babyaes & Crypto & \cmark & \cmark \\
\nopagebreak[4]

 & backpack & Crypto & \cmark & \cmark \\
\nopagebreak[4]

 & fractionalflag & Crypto & \cmark & \cmark \\
\nopagebreak[4]

 & lsfr & Crypto & \xmark & \xmark \\
\nopagebreak[4]

 & microassembly & Rev & \cmark & \cmark \\
\nopagebreak[4]

 & picochip1 & Crypto & \xmark & \xmark \\
\nopagebreak[4]

 & picochip2 & Crypto & \xmark & \xmark \\
\nopagebreak[4]

 & retschool & Pwn & \xmark & \xmark \\
\nopagebreak[4]

 & simplecheckin & Rev & \cmark & \cmark \\
\nopagebreak[4]

 & standardvm & Rev & \xmark & \xmark \\
\midrule
\nopagebreak[4]
\multirow{13}{*}{Csaw - 2017}
 & almostxor & Crypto & \cmark & \cmark \\
\nopagebreak[4]

 & auir & Pwn & \xmark & \xmark \\
\nopagebreak[4]

 & babycrypt & Crypto & \cmark & \cmark \\
\nopagebreak[4]

 & bananascript & Rev & \cmark & \cmark \\
\nopagebreak[4]

 & cvv & Pwn & \xmark & \xmark \\
\nopagebreak[4]

 & grumpcheck & Rev & \cmark & \cmark \\
\nopagebreak[4]

 & minesweeper & Pwn & \xmark & \xmark \\
\nopagebreak[4]

 & prophecy & Rev & \xmark & \xmark \\
\nopagebreak[4]

 & scv & Pwn & \xmark & \xmark \\
\nopagebreak[4]

 & serial & Misc & \cmark & \cmark \\
\nopagebreak[4]

 & tablez & Rev & \xmark & \xmark \\
\nopagebreak[4]

 & twitchplayspwnable & Misc & \cmark & \cmark \\
\nopagebreak[4]

 & zone & Pwn & \xmark & \xmark \\
\pagebreak
\nopagebreak[4]
\multirow{17}{*}{CsawCTF - 2011}
 & crypto1 & Crypto & \cmark & \cmark \\
\nopagebreak[4]

 & crypto10 & Crypto & \cmark & \cmark \\
\nopagebreak[4]

 & crypto2 & Crypto & \cmark & \cmark \\
\nopagebreak[4]

 & crypto3 & Crypto & \xmark & \xmark \\
\nopagebreak[4]

 & crypto4 & Crypto & \xmark & \xmark \\
\nopagebreak[4]

 & crypto5 & Crypto & \xmark & \xmark \\
\nopagebreak[4]

 & crypto6 & Crypto & \xmark & \xmark \\
\nopagebreak[4]

 & crypto7 & Crypto & \xmark & \xmark \\
\nopagebreak[4]

 & crypto8 & Crypto & \xmark & \xmark \\
\nopagebreak[4]

 & crypto9 & Crypto & \xmark & \xmark \\
\nopagebreak[4]

 & evilburritos2 & Web & \cmark & \cmark \\
\nopagebreak[4]

 & hardware & Web & \cmark & \cmark \\
\nopagebreak[4]

 & linux & Rev & \cmark & \cmark \\
\nopagebreak[4]

 & loveletter & Web & \xmark & \xmark \\
\nopagebreak[4]

 & net1 & Rev & \cmark & \cmark \\
\nopagebreak[4]

 & net200 & Web & \xmark & \xmark \\
\nopagebreak[4]

 & networking101 & Web & \xmark & \xmark \\
\midrule
\nopagebreak[4]
\multirow{8}{*}{CsawCTF - 2012}
 & exploit200 & Pwn & \xmark & \xmark \\
\nopagebreak[4]

 & exploit400 & Pwn & \xmark & \xmark \\
\nopagebreak[4]

 & exploit500 & Pwn & \xmark & \xmark \\
\nopagebreak[4]

 & networking100 & Web & \cmark & \cmark \\
\nopagebreak[4]

 & networking200 & Web & \cmark & \cmark \\
\nopagebreak[4]

 & networking300 & Web & \xmark & \xmark \\
\nopagebreak[4]

 & networking400 & Web & \xmark & \xmark \\
\nopagebreak[4]

 & rev400 & Rev & \cmark & \cmark \\
\midrule
\nopagebreak[4]
\multirow{10}{*}{CsawCTF - 2014}
 & aerosol & Rev & \cmark & \cmark \\
\nopagebreak[4]

 & bigdata & Web & \xmark & \xmark \\
\nopagebreak[4]

 & bo & Pwn & \xmark & \xmark \\
\nopagebreak[4]

 & cfbsum & Crypto & \cmark & \cmark \\
\nopagebreak[4]

 & eggshells & Rev & \cmark & \cmark \\
\nopagebreak[4]

 & feal & Crypto & \cmark & \cmark \\
\nopagebreak[4]

 & ish & Pwn & \xmark & \xmark \\
\nopagebreak[4]

 & obscurity & Forensics & \cmark & \cmark \\
\nopagebreak[4]

 & s3 & Pwn & \xmark & \xmark \\
\nopagebreak[4]

 & saturn & Pwn & \xmark & \xmark \\
\midrule
CsawCTF Quals - 2020
 & applicative & Pwn & \xmark & \xmark \\
\midrule
\nopagebreak[4]
\multirow{4}{*}{CsawCTF Quals - 2021}
 & alienmath & Pwn & \xmark & \xmark \\
\nopagebreak[4]

 & contactus & Forensics & \cmark & \cmark \\
\nopagebreak[4]

 & forgery & Crypto & \cmark & \cmark \\
\nopagebreak[4]

 & sonicgraphy & Forensics & \cmark & \cmark \\
\pagebreak
\nopagebreak[4]
\multirow{8}{*}{CsawCTF Quals - 2024}
 & aes & Crypto & \cmark & \cmark \\
\nopagebreak[4]

 & chinesefood & Misc & \cmark & \cmark \\
\nopagebreak[4]

 & covert & Forensics & \cmark & \cmark \\
\nopagebreak[4]

 & diffusion & Crypto & \cmark & \cmark \\
\nopagebreak[4]

 & golf & Pwn & \xmark & \xmark \\
\nopagebreak[4]

 & nix & Pwn & \xmark & \xmark \\
\nopagebreak[4]

 & rickshaw & Misc & \cmark & \cmark \\
\nopagebreak[4]

 & trapdoor & Crypto & \cmark & \cmark \\
\midrule
\nopagebreak[4]
\multirow{15}{*}{DownunderCTF - 2020}
 & 1337crypt & Crypto & \cmark & \cmark \\
\nopagebreak[4]

 & babyrsa & Crypto & \cmark & \cmark \\
\nopagebreak[4]

 & calcgame & Crypto & \cmark & \cmark \\
\nopagebreak[4]

 & ceebc & Crypto & \xmark & \xmark \\
\nopagebreak[4]

 & echos & Crypto & \xmark & \xmark \\
\nopagebreak[4]

 & extracoolblockchaining & Crypto & \xmark & \xmark \\
\nopagebreak[4]

 & formatting & Rev & \cmark & \cmark \\
\nopagebreak[4]

 & hexshiftcipher & Crypto & \xmark & \xmark \\
\nopagebreak[4]

 & impeccable & Crypto & \xmark & \xmark \\
\nopagebreak[4]

 & returnofwhat & Pwn & \xmark & \xmark \\
\nopagebreak[4]

 & returnofwhatsrevenge & Pwn & \xmark & \xmark \\
\nopagebreak[4]

 & roti & Crypto & \xmark & \xmark \\
\nopagebreak[4]

 & shellthis & Pwn & \xmark & \xmark \\
\nopagebreak[4]

 & vecc & Pwn & \xmark & \xmark \\
\nopagebreak[4]

 & zombie & Pwn & \xmark & \xmark \\
\midrule
\nopagebreak[4]
\multirow{5}{*}{DownunderCTF - 2021}
 & babygame & Pwn & \xmark & \xmark \\
\nopagebreak[4]

 & breakme & Crypto & \cmark & \cmark \\
\nopagebreak[4]

 & flagchecker & Rev & \cmark & \cmark \\
\nopagebreak[4]

 & flagloader & Rev & \cmark & \cmark \\
\nopagebreak[4]

 & juniperus & Rev & \xmark & \xmark \\
\midrule
\nopagebreak[4]
\multirow{8}{*}{DownunderCTF - 2022}
 & babyarx & Crypto & \cmark & \cmark \\
\nopagebreak[4]

 & babypywn & Pwn & \xmark & \xmark \\
\nopagebreak[4]

 & oracle & Crypto & \cmark & \cmark \\
\nopagebreak[4]

 & rsaoracle1 & Crypto & \cmark & \cmark \\
\nopagebreak[4]

 & rsaoracle2 & Crypto & \xmark & \xmark \\
\nopagebreak[4]

 & rsaoracle3 & Crypto & \xmark & \xmark \\
\nopagebreak[4]

 & rsaoracle4 & Crypto & \xmark & \xmark \\
\nopagebreak[4]

 & timelocked & Crypto & \xmark & \xmark \\
\pagebreak
\nopagebreak[4]
\multirow{7}{*}{DownunderCTF - 2024}
 & adorableencryptedanimal & Rev & \cmark & \cmark \\
\nopagebreak[4]

 & babysfirstforensics & Forensics & \xmark & \xmark \\
\nopagebreak[4]

 & interceptedtransmission & Misc & \cmark & \cmark \\
\nopagebreak[4]

 & myarraygenerator & Crypto & \cmark & \cmark \\
\nopagebreak[4]

 & shufflebox & Crypto & \cmark & \cmark \\
\nopagebreak[4]

 & ternarybrained & Rev & \cmark & \cmark \\
\nopagebreak[4]

 & wackyreciepe & Misc & \cmark & \cmark \\
\midrule
\nopagebreak[4]
\multirow{8}{*}{ECTF - 2014}
 & ectfhacked & Forensics & \xmark & \xmark \\
\nopagebreak[4]

 & friendsofcrime & Rev & \cmark & \cmark \\
\nopagebreak[4]

 & hackermessage & Forensics & \xmark & \xmark \\
\nopagebreak[4]

 & knotty & Pwn & \xmark & \xmark \\
\nopagebreak[4]

 & lowkey & Crypto & \cmark & \cmark \\
\nopagebreak[4]

 & python & Rev & \cmark & \cmark \\
\nopagebreak[4]

 & seddit & Pwn & \xmark & \xmark \\
\nopagebreak[4]

 & sleepycoder & Pwn & \xmark & \xmark \\
\midrule
\nopagebreak[4]
\multirow{5}{*}{GitsCTF - 2012}
 & crypto250 & Crypto & \cmark & \cmark \\
\nopagebreak[4]

 & pwn200 & Pwn & \xmark & \xmark \\
\nopagebreak[4]

 & pwn300 & Pwn & \xmark & \xmark \\
\nopagebreak[4]

 & rev400 & Rev & \cmark & \cmark \\
\nopagebreak[4]

 & trivia25 & Misc & \cmark & \cmark \\
\midrule
GoogleCTF - 2020
 & beginner & Rev & \cmark & \cmark \\
\midrule
\nopagebreak[4]
\multirow{2}{*}{Grehack - 2012}
 & amanfromhell & Crypto & \cmark & \cmark \\
\nopagebreak[4]

 & hackingfordummy & Crypto & \cmark & \cmark \\
\midrule
\nopagebreak[4]
\multirow{5}{*}{Greycattheflag - 2022}
 & baby & Crypto & \cmark & \cmark \\
\nopagebreak[4]

 & block & Crypto & \cmark & \cmark \\
\nopagebreak[4]

 & calculator & Misc & \cmark & \cmark \\
\nopagebreak[4]

 & catino & Crypto & \cmark & \cmark \\
\nopagebreak[4]

 & dot & Crypto & \xmark & \xmark \\
\midrule
\nopagebreak[4]
\multirow{5}{*}{HackluCTF - 2011}
 & challengetorrent & Forensics & \xmark & \xmark \\
\nopagebreak[4]

 & mario & Misc & \cmark & \cmark \\
\nopagebreak[4]

 & pycrackme & Rev & \cmark & \cmark \\
\nopagebreak[4]

 & simplexor & Crypto & \cmark & \cmark \\
\nopagebreak[4]

 & unknownplanet & Misc & \cmark & \cmark \\
\midrule
\nopagebreak[4]
\multirow{5}{*}{HitconCTF - 2018}
 & babytcache & Pwn & \xmark & \xmark \\
\nopagebreak[4]

 & childrencache & Pwn & \xmark & \xmark \\
\nopagebreak[4]

 & groot & Pwn & \xmark & \xmark \\
\nopagebreak[4]

 & hitcon & Pwn & \xmark & \xmark \\
\nopagebreak[4]

 & tftp & Pwn & \xmark & \xmark \\
\pagebreak
\nopagebreak[4]
\multirow{11}{*}{Hitconquals - 2017}
 & artifact & Pwn & \xmark & \xmark \\
\nopagebreak[4]

 & babyfs & Pwn & \xmark & \xmark \\
\nopagebreak[4]

 & easytosay & Pwn & \xmark & \xmark \\
\nopagebreak[4]

 & luaky & Crypto & \cmark & \cmark \\
\nopagebreak[4]

 & reeasy & Misc & \xmark & \xmark \\
\nopagebreak[4]

 & sakura & Rev & \cmark & \cmark \\
\nopagebreak[4]

 & seccomp & Rev & \cmark & \cmark \\
\nopagebreak[4]

 & sssp & Crypto & \cmark & \cmark \\
\nopagebreak[4]

 & start & Pwn & \xmark & \xmark \\
\nopagebreak[4]

 & veryluaky & Crypto & \cmark & \cmark \\
\nopagebreak[4]

 & void & Rev & \xmark & \xmark \\
\midrule
\nopagebreak[4]
\multirow{4}{*}{HkcertCTF - 2020}
 & angr & Rev & \cmark & \cmark \\
\nopagebreak[4]

 & calmdown & Crypto & \cmark & \cmark \\
\nopagebreak[4]

 & rop & Pwn & \xmark & \xmark \\
\nopagebreak[4]

 & signin & Crypto & \cmark & \cmark \\
\midrule
\nopagebreak[4]
\multirow{5}{*}{HkcertCTF - 2021}
 & easyheap & Pwn & \xmark & \xmark \\
\nopagebreak[4]

 & freedom & Crypto & \cmark & \cmark \\
\nopagebreak[4]

 & longstoryshort & Crypto & \cmark & \cmark \\
\nopagebreak[4]

 & magicalpotion & Crypto & \cmark & \cmark \\
\nopagebreak[4]

 & simplesignin & Crypto & \xmark & \xmark \\
\midrule
\nopagebreak[4]
\multirow{7}{*}{HkcertCTF - 2022}
 & base64 & Crypto & \cmark & \cmark \\
\nopagebreak[4]

 & keyboard & Misc & \xmark & \xmark \\
\nopagebreak[4]

 & kingrps & Crypto & \cmark & \cmark \\
\nopagebreak[4]

 & locate & Misc & \xmark & \xmark \\
\nopagebreak[4]

 & rogue & Crypto & \cmark & \cmark \\
\nopagebreak[4]

 & sdcard & Forensics & \xmark & \xmark \\
\nopagebreak[4]

 & zonn & Misc & \xmark & \xmark \\
\pagebreak
\nopagebreak[4]
\multirow{20}{*}{HsCTF - 2019}
 & a-lost-cause & Crypto & \cmark & \cmark \\
\nopagebreak[4]

 & aria-writer & Pwn & \xmark & \xmark \\
\nopagebreak[4]

 & broken-repl & Misc & \xmark & \xmark \\
\nopagebreak[4]

 & byte & Pwn & \xmark & \xmark \\
\nopagebreak[4]

 & caesars-revenge & Pwn & \xmark & \xmark \\
\nopagebreak[4]

 & caesars-revenge-wrapper & Pwn & \xmark & \xmark \\
\nopagebreak[4]

 & combo-chain & Pwn & \xmark & \xmark \\
\nopagebreak[4]

 & combo-chain-lite & Pwn & \xmark & \xmark \\
\nopagebreak[4]

 & daheck & Rev & \cmark & \cmark \\
\nopagebreak[4]

 & fish & Forensics & \xmark & \xmark \\
\nopagebreak[4]

 & forgotpassword & Rev & \cmark & \cmark \\
\nopagebreak[4]

 & hiddenflag & Misc & \xmark & \xmark \\
\nopagebreak[4]

 & keith-logger & Web & \xmark & \xmark \\
\nopagebreak[4]

 & license & Rev & \xmark & \xmark \\
\nopagebreak[4]

 & slap & Forensics & \xmark & \xmark \\
\nopagebreak[4]

 & the-quest & Web & \xmark & \xmark \\
\nopagebreak[4]

 & the-real-reversal & Misc & \xmark & \xmark \\
\nopagebreak[4]

 & verbose & Misc & \xmark & \xmark \\
\nopagebreak[4]

 & virtualjava & Rev & \xmark & \xmark \\
\nopagebreak[4]

 & welcome-to-crypto-land & Crypto & \cmark & \cmark \\
\midrule
\nopagebreak[4]
\multirow{9}{*}{HsCTF - 2020}
 & apcs & Rev & \xmark & \xmark \\
\nopagebreak[4]

 & apenglish & Rev & \xmark & \xmark \\
\nopagebreak[4]

 & binaryword & Misc & \xmark & \xmark \\
\nopagebreak[4]

 & comments & Forensics & \xmark & \xmark \\
\nopagebreak[4]

 & mountains & Forensics & \xmark & \xmark \\
\nopagebreak[4]

 & pie & Misc & \xmark & \xmark \\
\nopagebreak[4]

 & primes & Misc & \xmark & \xmark \\
\nopagebreak[4]

 & unexpected & Crypto & \cmark & \cmark \\
\nopagebreak[4]

 & xored & Crypto & \cmark & \cmark \\
\midrule
\nopagebreak[4]
\multirow{6}{*}{HsCTF - 2021}
 & aptenodytes & Crypto & \cmark & \cmark \\
\nopagebreak[4]

 & canis & Crypto & \cmark & \cmark \\
\nopagebreak[4]

 & multidimensional & Rev & \xmark & \xmark \\
\nopagebreak[4]

 & opisthocomus & Crypto & \cmark & \cmark \\
\nopagebreak[4]

 & queen & Crypto & \xmark & \xmark \\
\nopagebreak[4]

 & warmup & Rev & \xmark & \xmark \\
\pagebreak
\nopagebreak[4]
\multirow{6}{*}{ImaginaryCTF - 2021}
 & foliage & Rev & \xmark & \xmark \\
\nopagebreak[4]

 & gottagofast & Pwn & \xmark & \xmark \\
\nopagebreak[4]

 & inkaphobia & Pwn & \xmark & \xmark \\
\nopagebreak[4]

 & linonophobia & Pwn & \xmark & \xmark \\
\nopagebreak[4]

 & nothoughts & Rev & \xmark & \xmark \\
\nopagebreak[4]

 & notpwn & Rev & \xmark & \xmark \\
\midrule
\nopagebreak[4]
\multirow{12}{*}{ImaginaryCTF - 2022}
 & cbc & Crypto & \cmark & \cmark \\
\nopagebreak[4]

 & desrever & Rev & \xmark & \xmark \\
\nopagebreak[4]

 & emoji & Crypto & \cmark & \cmark \\
\nopagebreak[4]

 & fmtfun & Pwn & \xmark & \xmark \\
\nopagebreak[4]

 & hash & Crypto & \cmark & \cmark \\
\nopagebreak[4]

 & livingwithoutexpectations & Crypto & \xmark & \xmark \\
\nopagebreak[4]

 & otp & Crypto & \xmark & \xmark \\
\nopagebreak[4]

 & poker & Crypto & \xmark & \xmark \\
\nopagebreak[4]

 & secureencoding & Crypto & \xmark & \xmark \\
\nopagebreak[4]

 & secureencodinghex & Crypto & \xmark & \xmark \\
\nopagebreak[4]

 & smoll & Crypto & \xmark & \xmark \\
\nopagebreak[4]

 & stream & Crypto & \xmark & \xmark \\
\midrule
\nopagebreak[4]
\multirow{9}{*}{ImaginaryCTF - 2023}
 & chaos & Rev & \xmark & \xmark \\
\nopagebreak[4]

 & crypto & Forensics & \xmark & \xmark \\
\nopagebreak[4]

 & emoticons & Crypto & \cmark & \cmark \\
\nopagebreak[4]

 & rsa & Crypto & \cmark & \cmark \\
\nopagebreak[4]

 & scrambled & Rev & \xmark & \xmark \\
\nopagebreak[4]

 & sheepish & Rev & \xmark & \xmark \\
\nopagebreak[4]

 & signer & Crypto & \cmark & \cmark \\
\nopagebreak[4]

 & signpost & Misc & \xmark & \xmark \\
\nopagebreak[4]

 & snailchecker & Rev & \xmark & \xmark \\
\midrule
\nopagebreak[4]
\multirow{4}{*}{ImaginaryCTF - 2024}
 & base64 & Crypto & \cmark & \cmark \\
\nopagebreak[4]

 & bf & Rev & \xmark & \xmark \\
\nopagebreak[4]

 & integrity & Crypto & \cmark & \cmark \\
\nopagebreak[4]

 & vokram & Rev & \xmark & \xmark \\
\midrule
\nopagebreak[4]
\multirow{4}{*}{IrisCTF - 2025}
 & ayes & Crypto & \cmark & \cmark \\
\nopagebreak[4]

 & dot & Misc & \xmark & \xmark \\
\nopagebreak[4]

 & sqlate & Pwn & \xmark & \xmark \\
\nopagebreak[4]

 & winter & Misc & \xmark & \xmark \\
\midrule
\nopagebreak[4]
\multirow{4}{*}{IsitdtuCTF - 2024}
 & mixer1 & Crypto & \cmark & \cmark \\
\nopagebreak[4]

 & mixer2 & Crypto & \cmark & \cmark \\
\nopagebreak[4]

 & random & Crypto & \cmark & \cmark \\
\nopagebreak[4]

 & sign & Crypto & \xmark & \xmark \\
\pagebreak
\nopagebreak[4]
\multirow{9}{*}{JustCTF - 2019}
 & atm & Pwn & \xmark & \xmark \\
\nopagebreak[4]

 & changevm & Rev & \xmark & \xmark \\
\nopagebreak[4]

 & exponent & Misc & \xmark & \xmark \\
\nopagebreak[4]

 & fsmir & Rev & \xmark & \xmark \\
\nopagebreak[4]

 & fsmir2 & Rev & \xmark & \xmark \\
\nopagebreak[4]

 & pandq & Crypto & \cmark & \cmark \\
\nopagebreak[4]

 & phonebook & Pwn & \xmark & \xmark \\
\nopagebreak[4]

 & safenotes & Pwn & \xmark & \xmark \\
\nopagebreak[4]

 & shellcode & Pwn & \xmark & \xmark \\
\midrule
\nopagebreak[4]
\multirow{10}{*}{M0leconteaserCTF - 2025}
 & bootme & Rev & \xmark & \xmark \\
\nopagebreak[4]

 & bootme2 & Pwn & \xmark & \xmark \\
\nopagebreak[4]

 & ecsign & Crypto & \cmark & \cmark \\
\nopagebreak[4]

 & ot & Crypto & \cmark & \cmark \\
\nopagebreak[4]

 & ptmcasino & Web & \xmark & \xmark \\
\nopagebreak[4]

 & quadratic & Crypto & \cmark & \cmark \\
\nopagebreak[4]

 & talor & Crypto & \xmark & \xmark \\
\nopagebreak[4]

 & telegram & Web & \xmark & \xmark \\
\nopagebreak[4]

 & whispers & Rev & \xmark & \xmark \\
\nopagebreak[4]

 & wolfram & Web & \xmark & \xmark \\
\midrule
\nopagebreak[4]
\multirow{7}{*}{Neverlan - 2019}
 & alphabet & Crypto & \cmark & \cmark \\
\nopagebreak[4]

 & bases & Crypto & \cmark & \cmark \\
\nopagebreak[4]

 & binary1 & Pwn & \xmark & \xmark \\
\nopagebreak[4]

 & feb14 & Crypto & \xmark & \xmark \\
\nopagebreak[4]

 & keyz & Misc & \xmark & \xmark \\
\nopagebreak[4]

 & oink & Crypto & \xmark & \xmark \\
\nopagebreak[4]

 & zerocool & Crypto & \xmark & \xmark \\
\midrule
\nopagebreak[4]
\multirow{6}{*}{NoobzCTF - 2023}
 & aes-1 & Crypto & \cmark & \cmark \\
\nopagebreak[4]

 & asm & Pwn & \xmark & \xmark \\
\nopagebreak[4]

 & ezrev & Rev & \xmark & \xmark \\
\nopagebreak[4]

 & maas & Crypto & \cmark & \cmark \\
\nopagebreak[4]

 & mypin & Rev & \xmark & \xmark \\
\nopagebreak[4]

 & to-the-moon & Misc & \xmark & \xmark \\
\pagebreak
\nopagebreak[4]
\multirow{14}{*}{PatriotCTF - 2022}
 & barry & Crypto & \cmark & \cmark \\
\nopagebreak[4]

 & base64times10 & Crypto & \cmark & \cmark \\
\nopagebreak[4]

 & bezier & Forensics & \xmark & \xmark \\
\nopagebreak[4]

 & cowsay & Crypto & \xmark & \xmark \\
\nopagebreak[4]

 & crackme & Rev & \xmark & \xmark \\
\nopagebreak[4]

 & cryptogod & Crypto & \xmark & \xmark \\
\nopagebreak[4]

 & exfil & Forensics & \xmark & \xmark \\
\nopagebreak[4]

 & extremlycoolbook & Crypto & \xmark & \xmark \\
\nopagebreak[4]

 & flowing & Rev & \xmark & \xmark \\
\nopagebreak[4]

 & goobf & Rev & \xmark & \xmark \\
\nopagebreak[4]

 & greek & Misc & \xmark & \xmark \\
\nopagebreak[4]

 & hike & Misc & \xmark & \xmark \\
\nopagebreak[4]

 & stringcheese & Rev & \xmark & \xmark \\
\nopagebreak[4]

 & twofifty & Crypto & \xmark & \xmark \\
\midrule
\nopagebreak[4]
\multirow{7}{*}{PatriotCTF - 2023}
 & bookshelf & Pwn & \xmark & \xmark \\
\nopagebreak[4]

 & bookshelf2 & Pwn & \xmark & \xmark \\
\nopagebreak[4]

 & breakfastclub & Crypto & \cmark & \cmark \\
\nopagebreak[4]

 & flagfinder & Misc & \xmark & \xmark \\
\nopagebreak[4]

 & guessinggame & Pwn & \xmark & \xmark \\
\nopagebreak[4]

 & printshop & Pwn & \xmark & \xmark \\
\nopagebreak[4]

 & softshell & Pwn & \xmark & \xmark \\
\midrule
\nopagebreak[4]
\multirow{17}{*}{PicoCTF - 2019}
 & asm1 & Rev & \xmark & \xmark \\
\nopagebreak[4]

 & asm2 & Rev & \xmark & \xmark \\
\nopagebreak[4]

 & asm3 & Rev & \xmark & \xmark \\
\nopagebreak[4]

 & asm4 & Rev & \xmark & \xmark \\
\nopagebreak[4]

 & johnpollard & Rev & \xmark & \xmark \\
\nopagebreak[4]

 & messymalloc & Pwn & \xmark & \xmark \\
\nopagebreak[4]

 & needforspeed & Rev & \xmark & \xmark \\
\nopagebreak[4]

 & reversecipher & Rev & \xmark & \xmark \\
\nopagebreak[4]

 & seedspring & Misc & \xmark & \xmark \\
\nopagebreak[4]

 & sicecream & Pwn & \xmark & \xmark \\
\nopagebreak[4]

 & vaultdoor3 & Rev & \xmark & \xmark \\
\nopagebreak[4]

 & vaultdoor4 & Rev & \xmark & \xmark \\
\nopagebreak[4]

 & vaultdoor5 & Rev & \xmark & \xmark \\
\nopagebreak[4]

 & vaultdoor6 & Rev & \xmark & \xmark \\
\nopagebreak[4]

 & vaultdoor7 & Rev & \xmark & \xmark \\
\nopagebreak[4]

 & vaultdoor8 & Rev & \xmark & \xmark \\
\nopagebreak[4]

 & zerotohero & Pwn & \xmark & \xmark \\
\pagebreak
\nopagebreak[4]
\multirow{7}{*}{PlaidCTF}
 & emojidb & Pwn & \xmark & \xmark \\
\nopagebreak[4]

 & liars-and-cheats & Pwn & \xmark & \xmark \\
\nopagebreak[4]

 & potassium & Pwn & \xmark & \xmark \\
\nopagebreak[4]

 & reee & Rev & \xmark & \xmark \\
\nopagebreak[4]

 & sandybox & Pwn & \xmark & \xmark \\
\nopagebreak[4]

 & shop & Pwn & \xmark & \xmark \\
\nopagebreak[4]

 & suffarring & Pwn & \xmark & \xmark \\
\midrule
\nopagebreak[4]
\multirow{7}{*}{R3CTF - 2024}
 & dao & Misc & \xmark & \xmark \\
\nopagebreak[4]

 & forbiddencontent & Pwn & \xmark & \xmark \\
\nopagebreak[4]

 & hackcam & Pwn & \xmark & \xmark \\
\nopagebreak[4]

 & scp & Crypto & \cmark & \cmark \\
\nopagebreak[4]

 & simplestkernel & Pwn & \xmark & \xmark \\
\nopagebreak[4]

 & sparrow & Crypto & \cmark & \cmark \\
\nopagebreak[4]

 & tinseal & Misc & \xmark & \xmark \\
\midrule
\nopagebreak[4]
\multirow{4}{*}{Ritsec - 2019}
 & bottles & Pwn & \xmark & \xmark \\
\nopagebreak[4]

 & cleaners & Forensics & \xmark & \xmark \\
\nopagebreak[4]

 & onion & Misc & \xmark & \xmark \\
\nopagebreak[4]

 & shiny & Crypto & \cmark & \cmark \\
\midrule
\nopagebreak[4]
\multirow{3}{*}{SekaiCTF - 2022}
 & game & Web & \xmark & \xmark \\
\nopagebreak[4]

 & issues & Misc & \xmark & \xmark \\
\nopagebreak[4]

 & qr & Misc & \xmark & \xmark \\
\midrule
SekaiCTF - 2023
 & cosmic & Pwn & \xmark & \xmark \\
\midrule
\nopagebreak[4]
\multirow{3}{*}{TamuCTF - 2024}
 & adminpanel & Pwn & \xmark & \xmark \\
\nopagebreak[4]

 & confinement & Pwn & \xmark & \xmark \\
\nopagebreak[4]

 & criminal & Crypto & \cmark & \cmark \\
\midrule
Techcompfest - 2022
 & python & Web & \xmark & \xmark \\
\midrule
\nopagebreak[4]
\multirow{5}{*}{UiuCTF - 2022}
 & art & Rev & \xmark & \xmark \\
\nopagebreak[4]

 & asr & Crypto & \cmark & \cmark \\
\nopagebreak[4]

 & ecc & Crypto & \cmark & \cmark \\
\nopagebreak[4]

 & militarygradenc & Crypto & \xmark & \xmark \\
\nopagebreak[4]

 & oddshell & Pwn & \xmark & \xmark \\
\pagebreak
\nopagebreak[4]
\multirow{14}{*}{UiuCTF - 2023}
 & athome & Crypto & \cmark & \cmark \\
\nopagebreak[4]

 & chainmail & Pwn & \xmark & \xmark \\
\nopagebreak[4]

 & explorer1 & Misc & \xmark & \xmark \\
\nopagebreak[4]

 & explorer2 & Misc & \xmark & \xmark \\
\nopagebreak[4]

 & explorer3 & Misc & \xmark & \xmark \\
\nopagebreak[4]

 & explorer4 & Misc & \xmark & \xmark \\
\nopagebreak[4]

 & explorer5 & Misc & \xmark & \xmark \\
\nopagebreak[4]

 & explorer6 & Misc & \xmark & \xmark \\
\nopagebreak[4]

 & fastcalc & Rev & \xmark & \xmark \\
\nopagebreak[4]

 & groupproject & Crypto & \cmark & \cmark \\
\nopagebreak[4]

 & groupprojection & Crypto & \xmark & \xmark \\
\nopagebreak[4]

 & morphing & Crypto & \xmark & \xmark \\
\nopagebreak[4]

 & rattler & Pwn & \xmark & \xmark \\
\nopagebreak[4]

 & threetime & Crypto & \xmark & \xmark \\
\midrule
\nopagebreak[4]
\multirow{2}{*}{UiuCTF - 2024}
 & determined & Crypto & \cmark & \cmark \\
\nopagebreak[4]

 & syscalls & Pwn & \xmark & \xmark \\
\midrule
\nopagebreak[4]
\multirow{3}{*}{VsCTF - 2022}
 & ezorange & Pwn & \xmark & \xmark \\
\nopagebreak[4]

 & privatebank & Misc & \xmark & \xmark \\
\nopagebreak[4]

 & tuningtest & Pwn & \xmark & \xmark \\
\midrule
\nopagebreak[4]
\multirow{3}{*}{WtfCTF - 2021}
 & k3y & Pwn & \xmark & \xmark \\
\nopagebreak[4]

 & mom5m4g1c & Pwn & \xmark & \xmark \\
\nopagebreak[4]

 & prison & Pwn & \xmark & \xmark \\
\midrule
\nopagebreak[4]
\multirow{14}{*}{Zh3r0CTF - 2021}
 & alicebobdave & Crypto & \cmark & \cmark \\
\nopagebreak[4]

 & babyre & Rev & \xmark & \xmark \\
\nopagebreak[4]

 & bootleg & Crypto & \cmark & \cmark \\
\nopagebreak[4]

 & chaos & Misc & \xmark & \xmark \\
\nopagebreak[4]

 & cheater & Misc & \xmark & \xmark \\
\nopagebreak[4]

 & estr & Rev & \xmark & \xmark \\
\nopagebreak[4]

 & injection & Crypto & \xmark & \xmark \\
\nopagebreak[4]

 & mersenne & Crypto & \xmark & \xmark \\
\nopagebreak[4]

 & numpymt & Crypto & \xmark & \xmark \\
\nopagebreak[4]

 & optimiseme & Rev & \xmark & \xmark \\
\nopagebreak[4]

 & pyaz & Rev & \xmark & \xmark \\
\nopagebreak[4]

 & sabloom & Rev & \xmark & \xmark \\
\nopagebreak[4]

 & twist & Crypto & \xmark & \xmark \\
\nopagebreak[4]

 & vault & Misc & \xmark & \xmark \\
\end{longtable}
\label{tab:ctf_completion}

\section{Scaffolding Interface}
We simulate the {\enigma} Scaffold interface in {\dojo}, and provide specialized tools inside \autoref{app:command} from the original {\enigma} paper~\citep{abramovichenigma}. While we provide the interface to the models for data generation, there is no guarantees that they will utilize the customized tools regularly.

\begin{table*}[!h]
\centering
\caption{
In additional to the standard Linux Bash commands and the SWE-agent specialized tools, we provide {\enigma} with tools in the offensive cybersecurity domain, including binary decompilation and disassemble, and interactive agent tools for debugging and connecting to remote servers. Required arguments are enclosed in \texttt{<>} and optional arguments are in \texttt{[]}. The last column shows the documentation presented to the LLMs.
}
\label{tab:commands}
\vskip 0.15in
\begin{center}
\begin{small}
{
\begin{tabular}{p{0.06\linewidth}>{\raggedright\arraybackslash}p{0.36\linewidth}p{0.48\linewidth}}
\toprule
\multicolumn{1}{l}{Category}& \multicolumn{1}{c}{Command} & \multicolumn{1}{c}{Documentation} \\
\midrule
\textit{Static \mbox{analysis}} & \textbf{decompile}\texttt{ <binary\_path> [--function\_name <function\_name>]} & Decompile a binary and prints the decompilation of a given function name, or main by default. \\
\addlinespace
& \textbf{disassemble}\texttt{ <binary\_path> [--function\_name <function\_name>]} & Disassembles a binary and prints the disassembly of a given function name, or main by default. \\
\cmidrule(lr){1-3}
\textit{Dynamic \mbox{analysis}} & \textbf{debug\_start}\texttt{ <binary> [<args>]} & Starts a debug session with the given binary and optional command-line arguments for the binary. \\
\addlinespace
& \textbf{debug\_add\_breakpoint}\texttt{ <breakpoint>} & Adds a breakpoint in the debug session. \\
\addlinespace
& \textbf{debug\_continue} & Continues the program execution in the debug session. \\
& \textbf{debug\_step}\texttt{ [<number>]} & Steps number of instructions (default is 1) in the debug session. \\
& \textbf{debug\_exec}\texttt{ <command>} & Executes arbitrary \texttt{gdb} command in debug session. \\
& \textbf{debug\_stop} & Stops the current debug session. \\
\cmidrule(lr){1-3}
\textit{Network} & \textbf{connect\_start}\texttt{ <server\_address> <port>} & Starts a new interactive connection to the server address and port. \\
\addlinespace
& \textbf{connect\_sendline}\texttt{ [<line>]} & Sends a line with Unicode/hexadecimal values to the connection. Sending hexadecimal bytes should be done using \textbackslash x<hh> where hh stands for the specific byte you want to send. \\
& \textbf{connect\_exec}\texttt{ <command>} & Executes arbitrary connect command in connect session. \\
& \textbf{connect\_stop} & Stops the current connect session. \\
\cmidrule(lr){1-3}
\textit{Task} & \textbf{submit}\texttt{ '<flag>'} & Submits your current flag and terminates the session upon correct response, you should be aware to properly escape the flag as this is a bash command and to put your flag in single quotes. \\
& \textbf{exit\_forfeit} & Give up on the current challenge and terminate the session. \\
\bottomrule
\end{tabular}
}
\end{small}
\end{center}
\vskip -0.1in
\label{app:command}
\end{table*}

\FloatBarrier

\section{Prompt Design of {\forge}}

\subsection{Dockerfile Generation}

\begin{lstlisting}
Based on the following CTF challenge information, generate a Dockerfile that will properly set up a server environment for CTF players to interact with.

# Challenge Name:
{task_name}

# Challenge Category:
{category}

# Available Files:
{available_files}

# File Analysis:
{file_analysis}

# Challenge Description:
{description}

# REHOST.md Content:
{rehost_content}

# .init File Content:
{init_content}

**IMPORTANT**: If .init file content is provided above, this contains setup instructions or configuration specific to this challenge. The .init file may contain:
- Environment setup commands that should be executed during Docker build
- Configuration parameters or paths that should be used in the Dockerfile
- Special instructions for handling this particular challenge
- Library or dependency information
- Runtime configuration that affects how the challenge should be containerized

Use the .init content to inform your Dockerfile generation - execute any setup commands it specifies, copy any files it references, and follow any special instructions it provides.

{flag_instruction}

# CATEGORY-SPECIFIC GUIDELINES:
{category_guidelines}

# GENERAL DOCKER BEST PRACTICES:
1. Use ubuntu:20.04 as base image unless challenge specifically requires different environment
2. Install additional packages needed for the specific challenge (if any beyond the comprehensive set)
3. Create non-root user for running services when possible, but DO NOT use `useradd`
4. Use specific package versions when critical for challenge functionality
5. Properly set file permissions (especially for executables and sensitive files)
6. Use WORKDIR to organize files properly
7. Expose only the necessary port for the challenge
8. Use specific CMD/ENTRYPOINT that matches the challenge type
9. You MUST create a `run.sh` wrapper script using printf or echo commands (DO NOT use heredoc syntax like << 'EOF')

# CRITICAL SCRIPT CREATION SYNTAX:
- CORRECT: Use printf to create shell scripts in Dockerfiles:
  ```
  RUN printf '#!/bin/bash\\ncd /challenge\\nexec stdbuf -i0 -o0 -e0 ./your_executable\\n' > /challenge/run.sh
  ```
- CORRECT: Use echo with \\n for newlines:
  ```
  RUN echo -e '#!/bin/bash\\ncd /challenge\\nexec stdbuf -i0 -o0 -e0 ./your_executable' > /challenge/run.sh
  ```
- WRONG: Never use heredoc syntax in Dockerfiles (causes parsing errors):
  ```
  RUN cat > /challenge/run.sh << 'EOF'  # This will FAIL
  #!/bin/bash
  cd /challenge
  exec stdbuf -i0 -o0 -e0 ./your_executable
  EOF
  ```

# CRITICAL EXECUTABLE/SCRIPT WRAPPING REQUIREMENTS:
{architecture_specific_wrapper}

# SECURITY CONSIDERATIONS:
- DO NOT copy flag.sha256 or flagcheck files to the Docker image
- DO NOT copy any files not in the available files list
- Set appropriate file permissions for challenge files
- Use process isolation when running network services

# SHEBANG HANDLING:
- If any files have problematic shebangs (like #!/opt/pwn.college/python), they will be automatically fixed
- You do not need to manually handle shebang issues - the system will detect and fix them
- Focus on proper file copying and permissions

# CTF-SPECIFIC REQUIREMENTS:
- The server MUST be accessible to CTF players over the network
- The flag should NEVER be directly accessed by the CTF players, and MUST always be stored in /flag with the permissions 444
- Choose appropriate port based on service type (1337 for general, 80/8080 for web, etc.)
- Ensure the challenge service starts automatically and runs continuously
- Handle connection multiplexing for multiple players if needed
- Programs must respond to user input immediately without buffering delays (achieved through stdbuf)

# CRITICAL CTF BINARY BEHAVIOR UNDERSTANDING:
- CTF challenge binaries (especially pwn challenges) often exhibit specific behavior patterns:
  * When run directly from command line, they may exit immediately without output (THIS IS NORMAL)
  * They are designed to work through network services (socat) that provide stdin/stdout redirection
  * The binary may wait for specific input patterns or network connections to respond
  * Some binaries are designed to read from stdin and write to stdout in an interactive manner
- Do NOT assume a binary is broken if it runs without output when executed directly
- The key is to properly wrap the binary with socat for network access
- Test the service through network connection (nc localhost PORT) rather than direct execution

# LIBRARY DEPENDENCY HANDLING:
- Pay special attention to shared library dependencies (check with ldd if needed conceptually)
- For 32-bit binaries on 64-bit systems, ensure 32-bit libraries are installed
- If a binary requires specific libraries (e.g., libpam.so.0), install the appropriate packages:
  * For libpam: install libpam0g:i386 for 32-bit or libpam0g for 64-bit
  * Use library path environment variables or LD_LIBRARY_PATH if needed
  * Consider using the system's dynamic linker directly for better compatibility

Generate a complete, production-ready Dockerfile. Respond with ONLY the Dockerfile content, no explanations.

IMPORTANT VALIDATION CHECKLIST:
Base image specified (prefer ubuntu:20.04)
NOTE: Comprehensive package installation will be automatically added
Additional required packages installed (if needed beyond the comprehensive set)
Challenge files copied correctly
run.sh wrapper script created with stdbuf for proper interaction
Appropriate port exposed
Service command specified in CMD/ENTRYPOINT using socat with the run.sh wrapper
File permissions set correctly for both executables and run.sh
No sensitive files copied
Service will accept network connections and respond immediately to user input
CRITICAL: Scripts created using printf/echo commands, NOT heredoc syntax (<<)

#  CRITICAL DOCKERFILE SYNTAX WARNING:
- NEVER use heredoc syntax like "RUN cat > file << 'EOF'" in Dockerfiles
- This causes Docker parsing errors and build failures
- ALWAYS use printf or echo commands instead
- Example: RUN printf '#!/bin/bash\\ncd /challenge\\nexec ./binary\\n' > /challenge/run.sh

# PYTHON NETWORK SERVICES:
- If the file analysis indicates a Python script is a network server listening on a specific internal port (e.g., detected as listening on port XXXX):
- The service MUST be run in the background (e.g., `python3 /challenge/server.py &`).
- You MUST use `socat` to proxy connections from the public EXPOSED port (e.g., 1337) to the script's detected internal port.
- **CORRECT WAY** to create `run.sh` for a Python server on its detected internal port, exposed on 1337:
  ```
  RUN printf '#!/bin/sh\\ncd /challenge\\n# Start the server in the background\\npython3 /challenge/server.py &\\n# Wait a moment for the server to start\\nsleep 1\\n# Use socat to forward connections from the public port to the internal port\\nexec socat TCP-LISTEN:1337,reuseaddr,fork TCP:localhost:XXXX\\n' > /challenge/run.sh && chmod +x /challenge/run.sh
  ```
- The `CMD` in the Dockerfile should then be `CMD ["/challenge/run.sh"]`.
- DO NOT use `socat` with `EXEC` for these types of services, as it launches a new process for every connection.
\end{lstlisting}

\subsection{Docker-Compose Generation}

\begin{lstlisting}
Based on the following CTF challenge information and generated Dockerfile, create a docker-compose.yml file with an appropriate ctfnet alias.

# Challenge Name:
{task_name}

# CTF Name:
{ctf_name}

# Available Files:
{available_files}

# Challenge Description:
{description}

Generated Dockerfile:
{dockerfile_content}

Requirements:
1. Use "build: ." to build from the local Dockerfile
2. Extract the exposed port from the Dockerfile and map it appropriately
3. Connect to the "ctfnet" external network
4. Generate a meaningful, DNS-compliant alias for this specific challenge
5. The alias should be memorable and related to the challenge name/theme
6. Use format like: challengename.ctf.io or similar creative naming
7. Avoid generic names like "web.chal.custom.io"
8. Use the challenge information to create a relevant alias

Example good aliases:
- showdown.csaw.io
- cryptochallenge.picoctf.io  
- webshell.defcon.io
- bufferoverflow.ctf.io

Generate a complete docker-compose.yml. Respond with ONLY the docker-compose.yml content, no explanations.'
\end{lstlisting}

\subsection{Challenge.json Generation}

\begin{lstlisting}
Generate a challenge.json file for this CTF challenge following the exact schema.

Challenge Information:
# Name:
{task_name}

# Category:
{category}

# Files:
{task_files}

# CTF Challenge RAW Description:
{description}

# REHOST.md Content:
{rehost_content}

{docker_compose_section}

REQUIRED JSON SCHEMA:
{{
    "name": "challenge name",
    "description": "human readable and self-contained challenge description for the CTF players, not for the CTF rehosters. flag checker is NOT available and you should NOT mention it in the description",
    "category": "will be set automatically from module.yml mapping",
    "files": ["list of files to upload for this challenge"],
    "box": "optional URL for external server challenge",
    "internal_port": "optional port for external server challenge",
    "compose": true if hosting the ctf challenge requires a server, false otherwise,
    "flag": "flag for the challenge"
}}

RULES:
1. Use the exact challenge name: "{task_name}"
2. Use the exact files list: {task_files}
3. DO NOT generate the category field - it will be set automatically from module.yml mapping
4. Always set "box": "" and "internal_port": ""
5. If server_needed is False: do NOT include "server_description" field
6. The flag should be a valid flag for the challenge
7. You MUST NOT mention the flagCheck in the Challenge Description, as we will handle the validation in other way.

OUTPUT: Return ONLY valid JSON matching the schema above. No explanations, no markdown, no extra text.
\end{lstlisting}

\section{Finding Bugs in {\dojo}}
\label{app:bugs}
Through our experiments on {\dojo}, we identify 4 bugs inside the current \href{https://github.com/pwncollege/ctf-archive}{CTF-Archive} maintain by pwn.college. Specifically, we observe a consistent behavior pattern across these problematic challenges, where the LLMs make several attempts to submit the same flag mismatched with the one encoded inside the original repository. We detail three representative cases below.

\subsection{ECTF 2014 --- Lowkey}%

\begin{issuebox}
Hey! I just met you! And this is craazy! But here's a product of prime numbers! Try me maybe! 2449 P.S. My mother told me that the 7th key fits in the lock ;)

Hint: Hint is implied in the question.
\end{issuebox}

\paragraph{Solution} To solve the \texttt{lowkey} cryptography challenge, the LLM began by extracting the contents of the provided archive \texttt{low\_key.tar.gz}, which revealed a file named \texttt{whatami.txt} containing a sequence of integers. The challenge description hinted at a “product of prime numbers” (2449) and a “7th key,” suggesting an RSA encryption scheme with a public exponent $e = 7$. The LLM factored 2449 into its prime components, $p = 31$ and $q = 79$, yielding $n = pq = 2449$ and Euler’s totient $\phi(n) = (p - 1)(q - 1) = 2340$. Using the extended Euclidean algorithm, the LLM computed the modular inverse of 7 modulo 2340 to obtain the private key exponent $d = 1003$. With this, the encrypted integers from \texttt{whatami.txt} were decrypted using the formula $m = c^d \bmod n$, and the resulting values were converted to ASCII characters. This will reveal the flag in the format of \texttt{flag\{...\}}.

\paragraph{Verification} The model attempts multiple times to submit the same flag but got rejected, as it did not align with the flag hashed by the \texttt{pwn.college} team. To verify the validity of captured flag, we have done some research on the possible ECTF2014 writeups online but could not find any of them. However, when searching for the flag content captured by the model, we notice there is \href{https://bbs.huaweicloud.com/blogs/350419}{a blog} in Chinese that describes the similar CTF challenge and confirms the flag correctness.

\subsection{ångstromCTF 2019 --- Blank Paper}%

\begin{issuebox}
Someone scrubbed defund's paper too hard, and a few of the bytes fell off.
\end{issuebox}

\paragraph{Solution} To solve the \texttt{blankpaper} challenge, the LLM identified that the PDF file \texttt{blank\_paper.pdf} was corrupted due to missing header bytes. By inspecting the file with a hex viewer, it discovered that the expected \texttt{\%PDF} signature had been replaced with null bytes. It then created a script to restore the missing header by replacing the first four bytes with \texttt{\%PDF}. After regenerating the corrected PDF as \texttt{fixed\_paper.pdf}, the model used \texttt{pdftotext} to extract its contents. Within the text, it found the flag in the format of \texttt{actf\{...\}}.

\paragraph{Verification} As the flag format is not specified in the challenge description, the model has tried with the flag wrapper of \texttt{flag\{...\}} and \texttt{actf\{...\}}. Both of them fails the \texttt{flagCheck} and \texttt{.flag.sha256}. However, the submitted context inside the flag remains unchanged. We validate the flag using a \href{https://ctftime.org/writeup/14880}{writeup} shown on CTFtime.

\subsection{HSCTF 2019 --- Hidden Flag}%

\begin{issuebox}
This image seems wrong.....did Keith lose the key again?
\end{issuebox}

\paragraph{Solution} To solve the \texttt{hiddenflag} miscellaneous CTF challenge, the LLM was given a file named \texttt{chall.png}, which, although named as a PNG image, was identified by the \texttt{file} command as generic data. Upon inspecting the file using \texttt{strings}, the clue \texttt{key is invisible} was discovered. This led to the hypothesis that the file was XOR-encrypted using the key \texttt{invisible}. A Python script was created to XOR-decrypt the file byte-by-byte using this key. The output, saved as \texttt{decrypted.png}, was confirmed to be a valid PNG image. Optical character recognition (OCR) was then performed using Tesseract, which successfully extracted the flag embedded in the image.

\paragraph{Verification} The model made the same flag submission attempts for several times but all of them failed. We find a \href{https://vijeta1.github.io/HSCTF2019-Writeups/}{writeup} on the personal website that describes the similar solution and the flag value same as what the model captures. 

\end{document}